\documentstyle[eqsecnum,prb,aps,epsf]{revtex}  
\begin{document}
\title{Conformal Field Theory Approach to the Kondo
Effect\cite{Zakopane}} \author{Ian Affleck}  \address{  Canadian
Institute for Advanced Research and  Physics Department,
University of British Columbia, Vancouver, BC, V6T~1Z1, Canada  }
 \maketitle  \abstract{Recently, a new approach, based on boundary
conformal field theory, has been applied to a variety of quantum
impurity problems in condensed matter and particle physics.  A
particularly enlightening example is the multi-channel Kondo
problem.  In this review some earlier approaches to the Kondo
problem are discussed, the needed material on boundary conformal
field theory is developed and then this new method is applied to
the multi-channel Kondo problem.  }\vskip1cm

\centerline{OUTLINE} \vskip.5cm \noindent I. Renormalization Group
and Fermi Liquid Approaches to the Kondo Effect

A) Introduction to The Kondo Effect

B) Renormalization Group Approach

C) Mapping to a One Dimensional Model

D) Fermi Liquid Approach at Low T \\ \\ II. Conformal Field Theory
(``Luttinger Liquid'')  Techniques: Separation of Charge and Spin
Degrees of Freedom, Current Algebra, ``Gluing Conditions'',
Finite-Size Spectrum \\ \\   III. Conformal Field Theory Approach
to the Kondo Effect: ``Completing the Square''

A) Leading Irrelevant Operator,  Specific Heat, Susceptibility,
Wilson Ratio, Resistivity at $T>0$ \\ \\   IV. Introduction to the
Multi-Channel Kondo Effect: Underscreening and Overscreening

A) Large-k Limit

B) Current Algebra Approach \\ \\  V. Boundary Conformal Field
Theory \\ \\ VI. Boundary Conformal Field Theory Results on the
Multi-Channel Kondo Effect:

A) Fusion and the Finite-Size Spectrum

B) Impurity Entropy

C) Boundary Green's Functions: Two-Point Functions, T=0 Resistivity

D) Four-Point Boundary Green's Functions, Spin-Density Green's
Function

E) Boundary Operator Content and Leading Irrelevant Operator:

Specific Heat, Susceptibility, Wilson Ratio, Resistivity at $T>0$
\newpage \section{Renormalization Group and Fermi Liquid Approaches
to the Kondo Effect} \subsection{Introduction  to the Kondo Effect}

Most mechanisms contributing to the resistivity of metals,
$\rho(T)$, give either $\rho(T)$ decreasing to $0 $,  as
$T\rightarrow 0$ (phonons or  electron-electron interactions), or
$\rho(T)\rightarrow$ constant, as $T\rightarrow 0$ (non-magnetric
impurities). However, metals containing magnetic impurities show a
$\rho(T)$ which  increases as $T \rightarrow 0$. This was explained
by Kondo\cite{Kondo} in 1964 using a simple Hamiltonian:
\begin{equation}
H=\sum_{\vec{k}\alpha}\psi^{\dagger\alpha}_{\vec{k}}
\psi_{\vec{k}\alpha}\epsilon(k)+\lambda\vec{S}\cdot\sum_{\vec{k}\vec{k'}}
\psi^{\dagger}_{\vec k} \frac{\vec{\sigma}}{2}\psi_{\vec{k'}}
\end{equation}
 where $\psi_{\vec{k}\alpha}$'s are conduction electron
annihilation operators, (of momentum  $\vec{k}$, spin $\alpha$) and
$\vec{S}$ represents the spin of the  magnetic impurity with
$$[S^a, S^b]=i\epsilon^{abc}S^c.$$ The interaction term represents
an impurity spin interacting with the electron spin at $\vec{x}=0$.

With the above Hamiltonian, the Born approximation  gives:
$\rho(T)\sim\lambda^2$, independent of $T$. The next order term has
a divergent coefficient at  $T=0$: \begin{equation}
\rho(T)\sim[\lambda+\nu\lambda^2 \ln \frac{D}{T}+...]^2
\end{equation} Here $D$ is the band-width, $\nu$ the density of
states. This result stimulated an enormous amount of theoretical
work. As Nozi\`eres put it, ``Theorists  `diverged' on their own,
leaving the experiment realities way behind''.\cite{Nozieres1} What
happens at low $T$, i.e. $T \sim
T_K=De^{-\frac{1}{\upsilon\lambda}}$? In that case the
$O(\lambda^2)$ term will be as big as the term of $O(\lambda)$.
What about the $O(\lambda^3)$ term? Such questions helped lead to
the development of the renormalization group needed  to understand
the problem.

In particle physics such a growth of a coupling constant at  low
energies explains quark confinement (1973) and ``asymptotic
freedom" at  $E\rightarrow\infty$. To solve these problems,
Wilson\cite{Wilson} developed a very powerful numerical
renormalization group approach. The Kondo model was also ``solved"
by the Bethe ansatz\cite{Andrei,Weigmann} which gives the specific
heat and  magnetization. Nozi\`eres,\cite{Nozieres1,Nozieres2}
following ideas of Anderson\cite{Anderson} and Wilson,\cite{Wilson}
developed a very simple,  and in a sense exact, picture of the low
$T$ behaviour. With A. Ludwig, I have generalized and reformulated
Nozi\`eres'
approach\cite{Affleck1,Affleck2,Affleck3,Affleck4,Ludwig1,Affleck5,Affleck6,Ludwig2}
 using recent results in  conformal field theory. The latter
approach is very general and can be applied to a number of other
problems including multi-channel and higher spin Kondo effect,
\cite{Affleck1,Affleck2,Affleck3,Affleck4,Ludwig1,Affleck5,Affleck6,Ludwig2}
two (or more)-impurity  Kondo effects,\cite{Affleck7,Affleck8}
impurity assisted tunneling,\cite{Ralph1}  impurities in
one-dimensional conductors (``quantum wires")\cite{Wong} or 1D
antiferromagnets,\cite{Eggert,Sorensen} baryon-monopole
interaction,\cite{Affleck9}. Some of these problems, including the
multi-channel Kondo effect, exhibit  non-Fermi liquid behaviour.
These are among the very few exactly solved  problems that  do this
(the others are 1D Luttinger liquids). It has been suggested that
this may be connected with exotic behaviour of  certain compounds,
including high-$T_c$ superconductors.\cite{Cox,Ruckenstein}

 \subsection{Renormalization Group} 
\begin{figure}
\epsfxsize=10 cm
\centerline{\epsffile{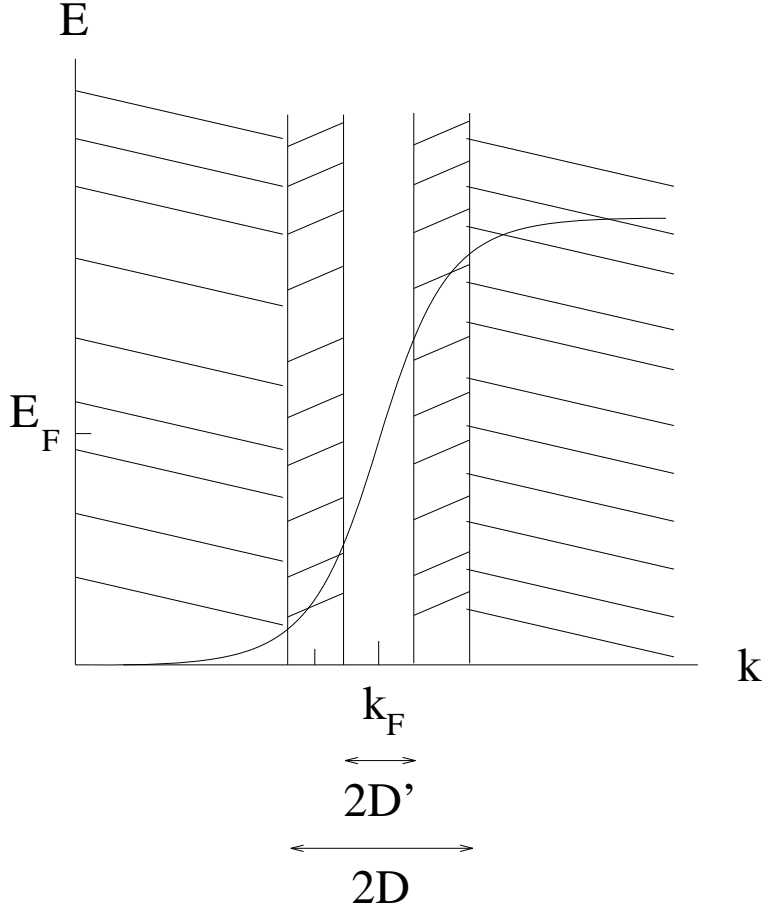}}
\caption{Reduction of the cut-off from $D$ to $D'$.}
\label{fig:red}
\end{figure}

We could integrate out
$\psi(k)$ for $k$ far from $k_{F}$, the Fermi wave-vector,
 and successively reduce the band-width $D$ to obtain a new
effective interaction.  [See Figure (\ref{fig:red}).] This is hard
to do exactly. At weak coupling one can do it perturbatively  in
$\lambda$. With the simplest approach, real-time, time-ordered
perturbation theory,  we expand $$T\exp
\left[-i\lambda\int\vec{S}(t)\cdot\psi^\dagger\frac{\vec{\sigma}}{2}
\psi(\vec{0},t)\right],$$ where the fields are in  the interaction
picture.

\begin{figure}
\epsfxsize=10 cm
\centerline{\epsffile{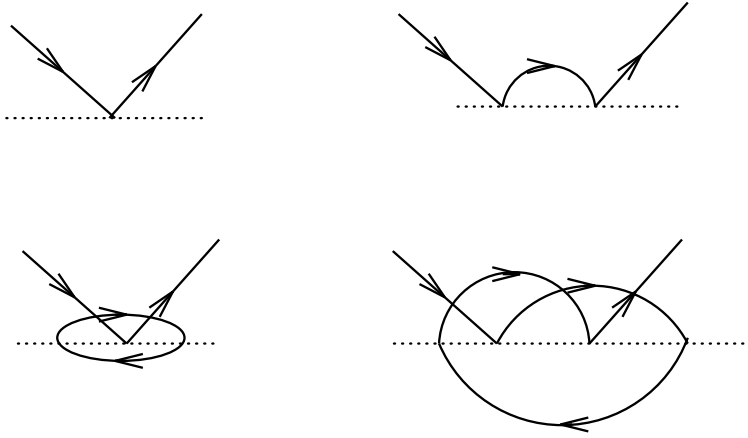}}
\caption{Feynman diagrams contributing to renormalization of the
Kondo coupling constant to third order.} \label{fig:pert}
\end{figure}

As $\vec{S}(t)$ is independent of $t$, we simply multiply powers of
$\vec{S}$ using
$$[S^a,S^b]=i\epsilon^{abc}S^c,~~~~\vec{S}^2=s(s+1).$$ We must
time-order $\vec{S}$'s which don't commute.  The first few diagrams
are shown in Figure (\ref{fig:pert}). In 2nd order in $\lambda$, we
have: $$-\frac{\lambda^2}{2}\int  dt\ dt'T(S^a(t)S^b(t'))\cdot
T[\psi^\dagger (t)\frac{\sigma^a}{2}\psi (t)\psi^\dagger
(t')\frac{\sigma^b}{2} \psi (t')],$$\\ which can be reduced, using
Wick's theorem, to: \begin{eqnarray} &&-\frac{1}{2}\lambda^2\int
dt\ dt'\psi^\dagger \left[
\frac{\sigma^a}{2},\frac{\sigma^b}{2}\right] \psi
T\langle\psi(t)\psi^\dagger (t')\rangle (\theta(t-t')S^a
S^b+\theta(t'-t)S^b S^a)\nonumber\\ =&&\frac{\lambda^2}{2}\int  dt\
dt'\psi^\dagger \frac{\vec{\sigma}}{2}\psi\cdot\vec{S}\
\hbox{sn}(t-t') \langle\psi(t)\psi^\dagger(t')\rangle
,\end{eqnarray} where sn $(t-t')$ is the sign-function which arises
{}from $T$-ordering  spins.

\begin{figure}
\epsfxsize=10 cm
\centerline{\epsffile{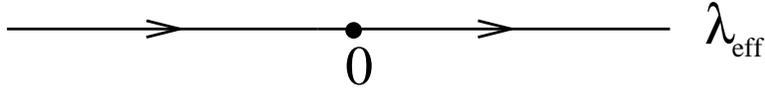}}
\caption{RG flow of the Kondo coupling.}
\label{fig:flowex}
\end{figure}

We see that the integral  \begin{equation}\int dt
\epsilon(t)G(t)=-i\int\frac{dt}{|t|}\end{equation} is divergent in
the infrared limit: $t\rightarrow\infty$, where
$G(t)=\langle\psi(t)\psi^\dagger(0)\rangle.$ But we only integrate
out electrons with $D'<k<D$ which gives $\ln D/D'$. To do it
explicitly we use the Fourier transformed form: \begin{eqnarray}
&&\int\frac{d^3k}{(2\pi)^3}\int\frac{d\omega }{2\pi}\left[
\frac{1}{i\omega +\delta} + \frac{1}{i\omega -\delta}\right]
\frac{i}{\omega-\epsilon_k+i\delta  \hbox{sn}(\epsilon_k)}\\ &&
=\int\frac{d^3\vec{k}}{(2\pi)^3}\frac{1}{|\epsilon_k|}\approx
2\nu\int  ^D_{D'}\frac{d\epsilon}{\epsilon}=2\nu\ln\frac{D}{D'}
.\end{eqnarray}
 Thus \begin{equation} \delta\lambda=\nu\lambda^2 \ln
\frac{D}{D'},\end{equation} and \begin{equation}
\frac{d\lambda}{d\ln D}=-\nu\lambda^2. \end{equation} We see that
lowering the band cut-off increases $\lambda$ or, defining a
length-dependent cut-off, $l\sim v_F/D$,
\begin{equation}\frac{d\lambda}{d\ln l}=\nu\lambda^2.\end{equation}
Integrating the equation (equivalent to performing an infinite sum
of diagrams), gives:
\begin{equation}\lambda_{\hbox{eff}}(D)=\frac{\lambda_0}{1-\nu\lambda_0
\ln \frac{D_0}{D}}.\end{equation} If $\lambda_0>0$
(antiferromagnetic), then $\lambda_{\hbox{eff}}(D)$ diverges at
$D\sim T_k \sim D_0e^{-\frac{1}{\nu\lambda_0}}$, If $\lambda_0<0$
(ferromagnetic), $\lambda_{\hbox{eff}}(D)\rightarrow 0$.  See
Figure (\ref{fig:flowex}).

The behaviour at temperature $T$ is  determined by
$\lambda_{\hbox{eff}}(T)$:  $\rho(T)\rightarrow 0$ as $T\rightarrow
0$ for the ferromagnetic case. What happens for the
antiferromagnetic case?

\subsection{Mapping to a One-Dimensional Model}

\begin{figure}
\epsfxsize=10 cm
\centerline{\epsffile{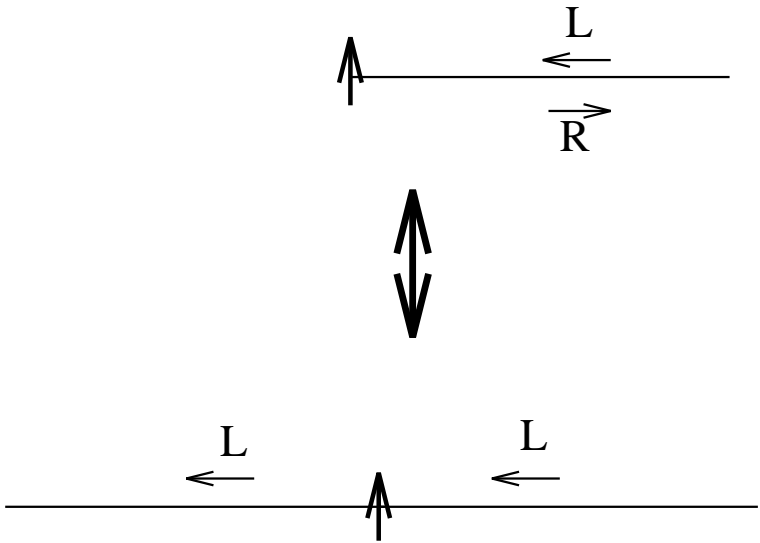}}
\caption{Reflecting the left-movers to the negative axis.}
\label{fig:1DLR}
\end{figure}

The above discussion  can be simplified if we map the model into a
one  dimensional one. We assume a spherically symmetric
$\epsilon(\vec k)$, \begin{equation}
\epsilon(k)=\frac{k^2}{2m}-\epsilon_F\approx v_F(k-k_F),
\end{equation}  and  a $\delta-$function Kondo  interaction. There
is only s-wave scattering, i.e.  \begin{eqnarray}
\psi(\vec{k})&=&\frac{1}{\sqrt{4\pi}k}\psi_0(k)+\mbox{higher
harmonics}, \nonumber\\ H_0&=&\int dk \psi_{0 k}^\dagger\psi_{0
k}\epsilon(k)+ \mbox{higher harmonics}, \nonumber\\
H_{\hbox{INT}}&=&\lambda v_F\nu\int dkdk'\psi_{0,k}^\dagger{\vec
\sigma \over 2}\psi_{0,k'} \cdot \vec S, \end{eqnarray} where
$\nu=k_F^2/2\pi^2v_F$ is the density of states per spin. This can
also  be written in terms of  radial co-ordinate. We eliminate all
modes except for a band width 2D: $|k-k_F|<D$. Defining left and
right movers (incoming and outgoing waves), \begin{equation}
\Psi_{L,R}(r)\equiv\int^\wedge_{-\wedge} dke^{\pm
ikr}~~\psi_0(k+k_F), ~~~\Rightarrow~~~  \psi_L(0)=\psi_R(0),
\end{equation} we have \begin{eqnarray}
H_0&=&\frac{v_F}{2\pi}\int^{\infty}_0 dr(\psi^\dagger
_Li\frac{d}{dr}\psi_L-\psi^ \dagger _Ri \frac{d}{dr}\psi_R)~~~
\mbox{(note the unconventional normalization)},\nonumber\\
H_{\hbox{INT}}&=&v_F\lambda\psi_L(0)^\dagger
\frac{\vec{\sigma}}{2}\psi_L(0)\cdot\vec{S}. \label{HINT}
\end{eqnarray} Here we have redefined a dimensionless Kondo
coupling, $\lambda \rightarrow \lambda \nu$. Using the notation
\begin{equation} \psi_L=\psi_L(x,\tau)=\psi_L(z=\tau+ix),~~
\psi_R(x,\tau)=\psi_R(z^*=\tau-ix) ,\end{equation}  where $\tau$ is
imaginary time and $x=r$, (and we set $v_F=1$) we have
\begin{equation}
\langle\psi_L(z)\psi_{L}^+(0)\rangle=\frac{1}{z},~~
\langle\psi_R(z^*)\psi_R^\dagger(0)\rangle=\frac{1}{z^*}.
\end{equation} Alternatively, since  \begin{equation}
\psi_L(0,\tau)=\psi_R(0,\tau)~~~~~\psi_L=\psi_L(z),~~\psi_R=\psi_R(z^*),
\end{equation} we may consider $\psi_R$ to be the continuation of
$\psi_L$ to the negative  $r$-axis:  \begin{equation} \psi_R(x,
\tau)\equiv\psi_L(-x,\tau). \end{equation} Now we obtain a
relativistic $(1+1)$ dimensional field theory ( a ``chiral'' one,
containing   left-movers only) interacting with the impurity at
$x=0$ with \begin{equation} H_0={v_F\over
2\pi}\int_{-\infty}^\infty dx\psi^\dagger_Li {d \over
dx}\psi_L\label{Hochi}\end{equation} and $H_{\hbox{INT}}$ as in Eq.
(\ref{HINT}).
 See Figure (\ref{fig:1DLR}).

\subsection{Fermi Liquid Approach at Low T} What is the
$T\rightarrow 0 $ behavior of the antiferromagetic Kondo model? The
simplest assumption is $\lambda_{\hbox{eff}}\rightarrow\infty$. But
what does that really mean? Consider the strong coupling limit of a
lattice model,\cite{Nozieres1} for convenience, in spatial
dimension $D=1$. ($D$ doesn't really matter since we can always
reduce the model to $D=1$.) \begin{equation}
H=t\sum_{i}(\psi^\dagger_i\psi_{i+1}+\psi^\dagger_{i+1}
\psi_i)+\lambda\vec{S}\cdot \psi^\dagger_0
\frac{\vec\sigma}{2}\psi_0\label{lattice} \end{equation} Consider
the limit $\lambda>>|t|$. The groundstate of the interaction term
will be the following configuration: one electron at the site $0$
forms a singlet with the impurity:
$|\Uparrow\downarrow\rangle-|\Downarrow\uparrow\rangle$.  (We
assume~~$S_{IMP}=1/2$). Now we do perturbation theory in $t$.  We
have the following low energy states: an arbitary electron
configuration occurs on all
 other sites-but  other electrons or holes are forbidden to enter
the site-$0$, since that  would destroy the singlet state, costing
an energy, $\Delta E \sim \lambda >>t$. Thus we simply form free
electron Bloch states with the boundary condition $\phi(0)=0$,
where $\phi(i)$ is the single-electron wave-function.  Note that at
zero Kondo coupling, the parity even single particle wave-functions
are of the form $\phi (i)=\cos ki$  and the parity odd ones are of
the form $\phi (i)=\sin ki$. On the other hand, at
$\lambda\rightarrow \infty$ the parity even wave-functions become
$\phi (i)=|\sin ki|$, while the parity odd ones are unaffected.

The behaviour of the parity even channel corresponds to a $\pi/2$
phase shift in  the s-wave channel. \begin{equation} \phi_j\sim
e^{-ik|j|}+e^{+2i\delta}e^{ik|j|}, ~~\delta=\pi/2. \end{equation}
In terms of left and right movers on $r>0$ we have changed the
boundary condition,  \begin{eqnarray} \psi_L(0)&=&\psi_R(0),
{}~~~~\lambda=0, \nonumber\\ \psi_L(0)&=&-\psi_R(0),~~~~
\lambda=\infty. \end{eqnarray} The strong coupling fixed point is
the same as the weak coupling  fixed point except for a change in
boundary conditions (and the removal of the  impurity). In terms of
the left-moving description of the $P$-even sector, the phase of
the left-mover  is shifted by $\pi$ as it passes the origin.
Imposing another boundary condition a distance $l$ away quantizes
$k$: \begin{eqnarray}
\psi(l)=\psi_L(l)+\psi_R(l)=\psi_L(l)+\psi_L(-l)&=&0,\nonumber\\
 \lambda&=&0:~~~~~~k=\frac{\pi}{l}(n+1/2)\nonumber\\
\lambda&=&\infty :~~~~~~k=\frac{\pi n}{l} \end{eqnarray}

\begin{figure}
\epsfxsize=10 cm
\centerline{\epsffile{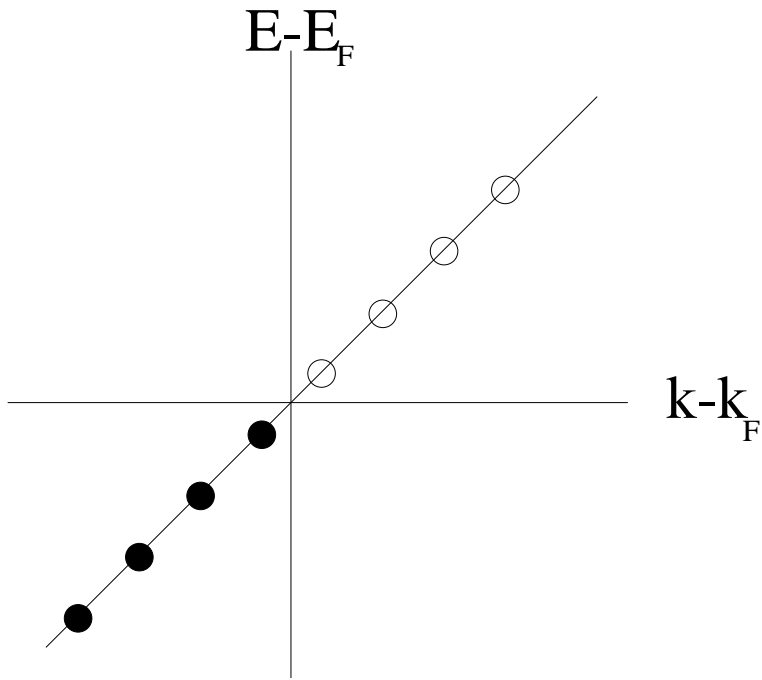}}
\caption{Free fermion energy levels with antiperiodic boundary
conditions.}  \label{fig:anti} \end{figure}

\begin{figure}
\epsfxsize=10 cm
\centerline{\epsffile{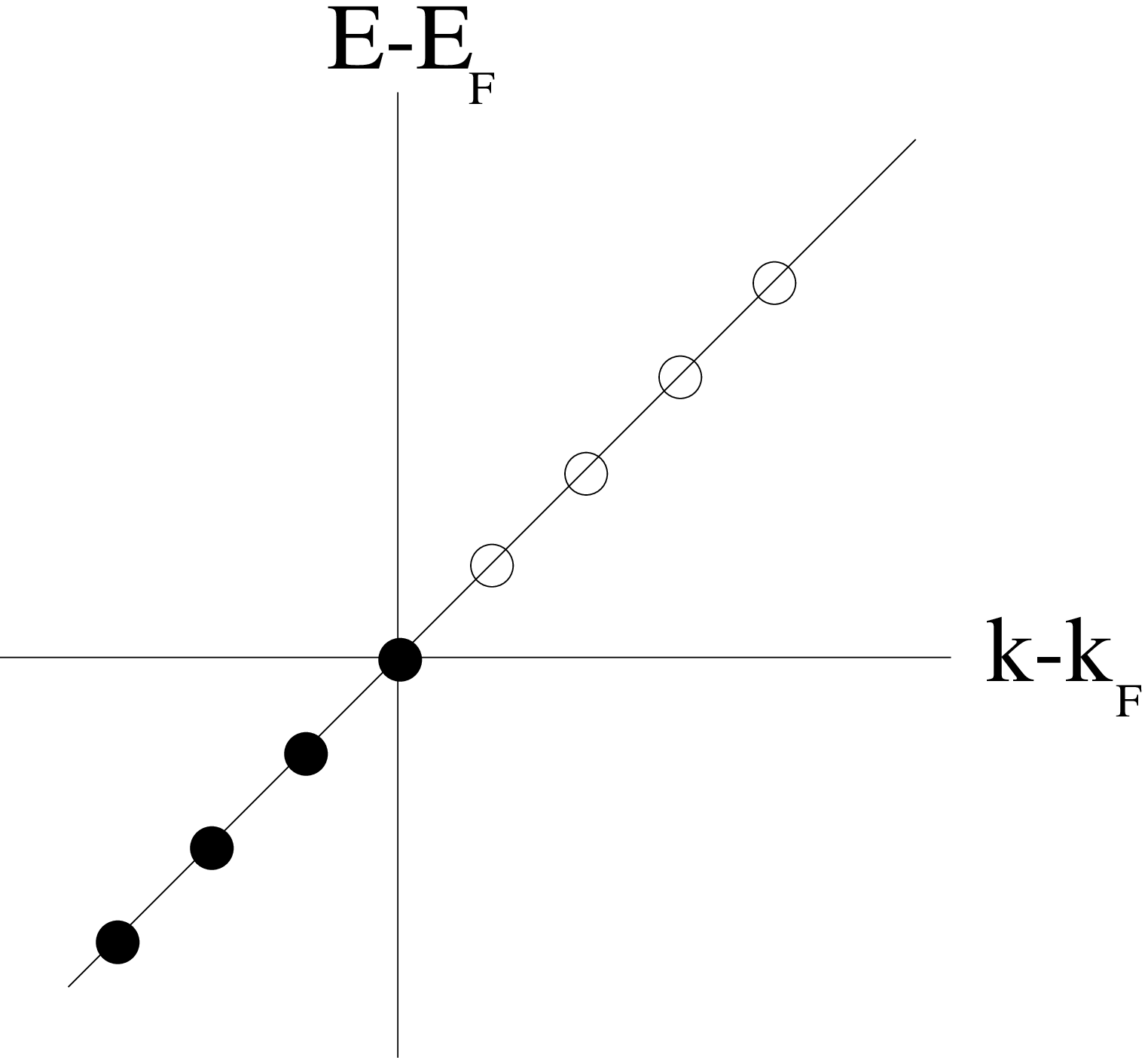}}
\caption{Free fermion energy levels with periodic
boundary conditions.}
\label{fig:per}
\end{figure}

\begin{figure}
\epsfxsize=10 cm
\centerline{\epsffile{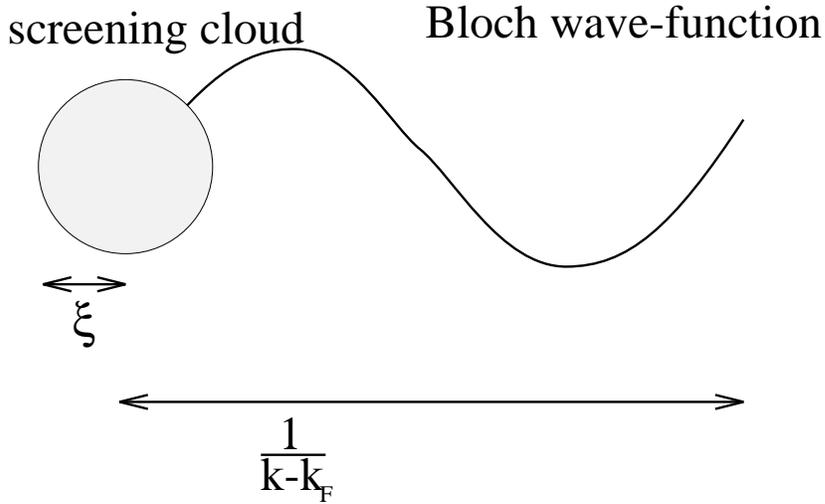}}
\caption{Non-interacting Bloch states with a vanishing
boundary condition occur for $|k-k_F|<<v_F/T_K$.}
\label{fig:cloud}
 \end{figure}

Near the Fermi surface the energies are linearly spaced. Assuming
particle-hole symmetry, the  Fermi energy lies midway between
levels or on a level.
  [See Figures (\ref{fig:anti}) and (\ref{fig:per}).] The two
situations switch with the phase shift. Wilson's numerical RG
scheme\cite{Wilson} involves  calculating the low-lying spectrum
numerically and looking for this shift. This indicates that
$\lambda$ renormalizes to $\infty$ even if it is  initially small.
However, now we expect the screening to take place over a longer
length  scale  \begin{equation} \xi
\sim\frac{v_F}{T_K}\sim\frac{v_F}{D}e^{1/\nu\lambda}
.\end{equation} In other words, the wave function of the screening
electron has this scale.  We get low energy  Bloch states of free
electrons only for $|k-k_F|<<1/\xi$ (so we must take $l>>\xi$).
[See Figure (\ref{fig:cloud}).] The free electron theory with a
phase shift corresponds to a universal stable low energy fixed
point for the Kondo problem.  This observation determines the $T=0$
resistivity for an array of Kondo impurities at random locations of
low density $n_i$.  It is the same as for non-magnetic s-wave
scatterers with a $\pi /2$ phase shift at the Fermi energy.
$\delta = \pi /2$ gives the so-called unitary limit resistivity:
\begin{equation} \rho_{\hbox{u}} = {3n_i\over \pi\nu^2
v_F^2e^2}.\end{equation}

The low-$T$ behaviour, so far, seems trivial. Much of the
interesting behaviour  comes from the leading irrelevant operator.
The impurity spin has disappeared (screened) from the description
of the low -$T$  physics. However certain interactions between
electrons are generated (at the  impurity site only) in the process
of eliminating the impurity spin.  We can determine these by simply
writing the lowest dimension operators allowed  by symmetry.

It is simplest to work in the 1D formulation, with left-movers only.
We write the interaction in terms of $\psi_L$, obeying the new
boundary condition  (but \underline{not} the impurity spin). The
dimension of the operator is determined as in 1D field theory
\begin{equation} H=\int dx\psi^\dagger_L i\frac{d}{dx}\psi_L +....
\end{equation} The length and time dimensions are equivalent (we
convert with $v_F$),  $$ [H]=E\Rightarrow[\psi]=E^{\frac{1}{2}}. $$
The interactions are local  $$ \delta H=\sum_{i}\lambda_iO_i(x=0),
{}~~~~[\lambda_i]+[O_i]=1.$$ So $\lambda_i$ has negative energy
dimension if $[O_i]>1$, implying that it is irrelevant. In RG
theory one usually defines a dimensionless coupling constant by
multiplying powers of the cut-off $D$,  if $$[\lambda_i]=E^{-a},
{}~~~~\tilde \lambda_i\equiv \lambda_iD^a,$$ $\tilde \lambda_i$
decreases as we lower $D$:  \begin{equation} \frac{d\tilde
\lambda_i}{dlnD}=a\tilde \lambda_i. \end{equation} Such a coupling,
with $a>0$, produces no infrared divergences in perturbation
theory.  The ultraviolet ones are cancelled by the expicit factors
of the ultraviolet cut-off, $D$, appearing in the Lagrangian,
$\tilde \lambda O/D^a$.
 What are the lowest dimension operators allowed by  symmetry?
Consider $\psi^{+\alpha}(0)\psi_\alpha(0)$. This has $d=1$.
However, it is not allowed because it breaks particle-hole
symmetry. If particle-hole symmetry is broken  then we do get this,
a potential scattering term; it adds a term to the phase shift.
Consider another term, \begin{equation} i\psi^{\dagger
\alpha}\frac{d}{dx}\psi_\alpha (0)- i{d\over dx}\psi^{\dagger
\alpha}\psi_\alpha (0).\end{equation}    This has d=2 .  This term
produces a $k$-dependent phase shift. The only other term with
$d\leq 2$ is  $\psi^{\dagger \uparrow}\psi_{\uparrow}\psi^{\dagger
\downarrow}\psi_{\downarrow} $. This term represents the
electron-electron interaction induced by an impurity  spin-flip.
The first electron flips the impurity spin. This makes it possible
for the second electron to flip it back if the electron spin is
correct. These are the only $d\leq 2$ operators.  There are no
relevant ($d\leq 1$) operators, implying the stability of the low
energy fixed point.  Note that, by contrast, the high energy, zero
Kondo coupling, fixed point is unstable.  The dimension 1 operator
(the Kondo interaction) can occur there because of the presence of
the impurity spin.

\begin{figure}
\epsfxsize=10 cm
\centerline{\epsffile{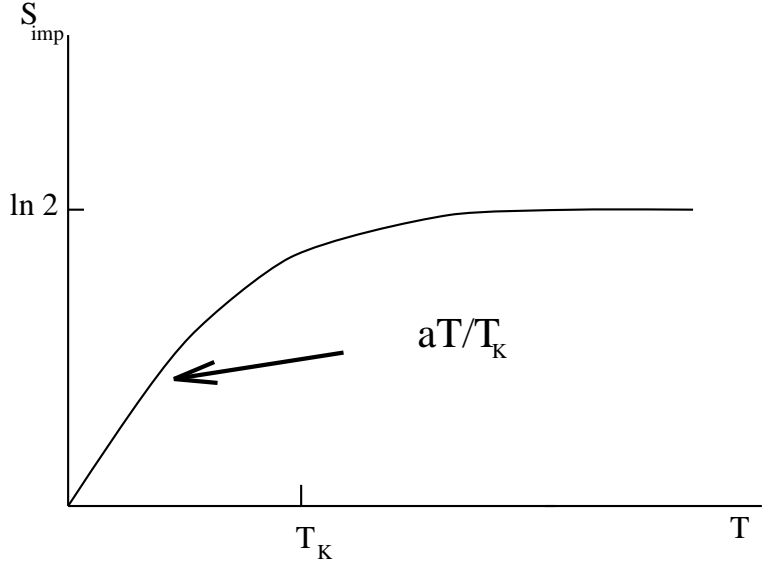}}
\caption{Qualitative behaviour of the impurity entropy.}
\label{fig:entropy} \end{figure}

 We can't calculate these two coupling constants exactly except by
using  complicated methods: Wilson's numerical melthod or the Bethe
ansatz. They both have dimension $E^{-1}$. We expect them to be
$O[1/{T_K}]$
 by a standard scaling argument.  That is, functions of the cut-off
$D$ and coupling constant $\lambda$ can be replaced by functions of
the reduced cut-off, $D'$ and the renormalized coupling constant,
$\lambda_{\hbox{eff}}(D')$:
$f[D,\lambda]=f[D',\lambda_{\hbox{eff}}(D')]$. We can lower the
cut-off down to $T_K$ where $\lambda_{\hbox{eff}}$ is $O(1)$ so
$f=f(T_K,1)=f(T_K)$.  This is a characteristic scale  introduced by
the infrared divergences of perturbation theory. For $T_K<<D
(\lambda<<1)$, Nozi\`eres argued that the two irrelevant coupling
constants have a universal ratio.  So there is only one  unknown
parameter (``the Wilson number''). Essentially all low-temperature
information is given by this irrelevant  coupling constant, if it
was not already determined by the $\pi/2$ phase shift  at $k_F$.  I
will give a different derivation of the ratio of  the two coupling
constants later using conformal field theory.  We now simply do
perturbation theory in the irrelevant coupling  constant
$\sim1/{T_K}$. We can determine powers of $T$ by dimensional
analysis. For the specific heat we find: \begin{equation}  C \sim
\frac{\pi}{3v_F}lT+a\frac{T}{T_K}.\end{equation}  This is the
specific heat for the one-dimensional system with a single impurity
at the origin.  Note that the first term is simply the specific
heat of the free system, proportional to system length.  The second
term is  independent of length and is the impurity specific heat.
It is the result of first order perturbation theory in the
irrelevant coupling constant, of $O(1/T_K)$.
 The (linear) power of $T$ can be fixed by dimensional analysis;
$a$ is a pure number.  Note that while this is formally an
``irrelevant'' contribution, it in fact gives the leading impurity
specific heat at low $T$.  To obtain the specific heat for the
three dimensional system we simply multiply the first term by the
ratio $\nu V/(l/2\pi v_F)$ i.e. the ratio of densities of states
per unit energy.  For a dilute random array the last term gets
multiplied by the number of impurities. At high T we get,
approximately, the entropy for a decoupled s=1/2 impurity:
\begin{equation} S(T) = {\pi l\over 3v_F}T + \ln 2.\end{equation}
At low $T$,  the impurity entropy decreases to 0: \begin{equation}
S(T) = {\pi l\over 3v_F}T+ {aT\over T_K}.\end{equation}  In general
we may write: \begin{equation} S(T) - {\pi l\over 3v_F}T \equiv
S_{\hbox{imp}}= g(T/T_K),\end{equation} where $g$ is a  scaling
function which is universal for weak bare coupling.  See Figure
(\ref{fig:entropy}). The behaviour of $g(x)$ for small arguments is
determined by RG-improved weak coupling perturbation theory.   It's
behaviour at low T is determined from the theory of the low energy
fixed point.  Its behaviour at arbitrary $T/T_K$ is a property of
the universal crossover between fixed points.  It has been found
{}from the Bethe ansatz.

  Similarly, the susceptibility, at $T=0$, is given by:
\begin{equation}  \chi \sim \frac{l}{2\pi v_F}+  \frac{b}{T_K}.
\end{equation} The ratio $b/a$ is universal since the coupling
constant ($1/T_K$) drops  out.  This is known as the Wilson Ratio.

\begin{figure}
\epsfxsize=10 cm
\centerline{\epsffile{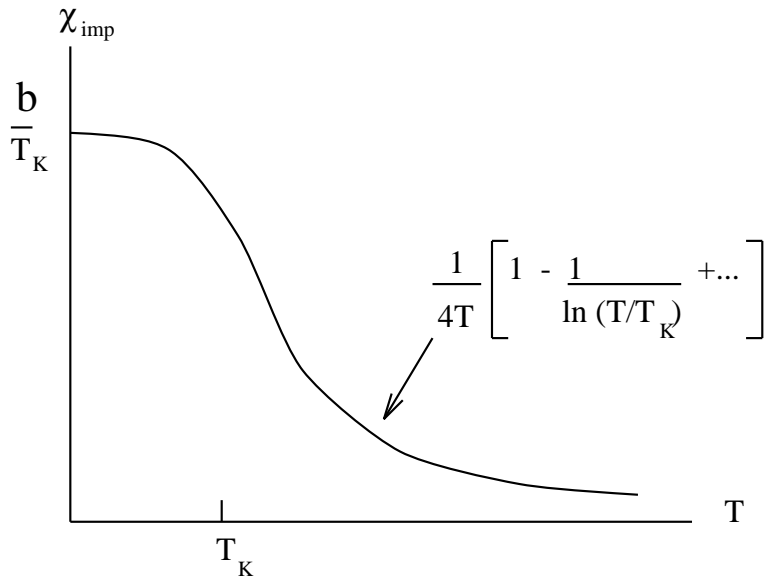}}
\caption{Qualitative behaviour of the impurity susceptibility.}
\label{fig:chi}
\end{figure}

At high $T$, we must (for weak bare coupling) obtain approximately
the results for a free spin: \begin{equation} \chi  \sim {l\over
2\pi v_F} + {1\over 4T}.\end{equation}  At lower $T$, using RG
improved perturbation theory this becomes: \begin{equation} \chi
\sim {l\over 2\pi v_F} + {1\over 4T}\left[1-{1\over \ln (T/T_K)}+
...\right].\end{equation} In general, we may write:
\begin{equation} \chi - {l\over 2\pi v_F} \equiv \chi_{\hbox{imp}}=
{1\over T}f(T/T_K),\end{equation} where $f(T/T_K)$ is another
universal scaling function.  See Figure (\ref{fig:chi}).

 The temperature dependent part of the low $T$  resistivity for the
dilute random array is $2nd$ order in perturbation  theory,
\begin{equation} \rho=\rho_{\hbox{u}}[1-d\left(
\frac{T}{T_K}\right)^2], \end{equation} where $d$ is another
dimensionless constant.  The second term comes from second order
perturbation theory in the irrelevant coupling constant. Another
universal ratio can be formed.  I will discuss this in Sec. III, in
some detail, using the CFT approach.

\begin{figure}
\epsfxsize=10 cm
\centerline{\epsffile{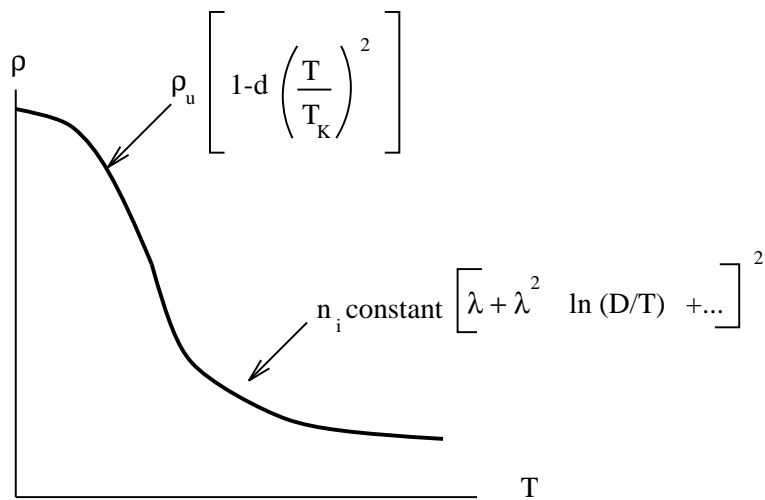}}
\caption{Qualitative behaviour of the resistivity.}
\label{fig:resex}
\end{figure}

We now expect a scaling behaviour: $$\rho (T) = n_if(T/T_K),$$
sketched in Figure (\ref{fig:resex}).

\section{Conformal Field Theory Techniques} It is very useful to
bosonize free fermions to understand the Kondo effect. This allows
separation of spin and  charge degrees of freedom which  greatly
simplifies the problem.

We start by considering a left-moving spinless fermion field with
Hamiltonian density: \begin{equation} {\cal
H}=\frac{1}{2\pi}\psi_L^{\dagger}i\frac{d}{dx}\psi_L.
\end{equation}

Define the current (=density) operator, \begin{eqnarray} J_L(x-t)
&=&:\psi_L^+\psi_L:(x,t)\nonumber\\
       &=&\lim_{\epsilon\rightarrow  0}[\psi_L(x)\psi_L
(x+\epsilon)-\langle 0|\psi_L(x)\psi_L(x+\epsilon)|0\rangle ]
\end{eqnarray} (Henceforth we generally drop the subscripts ``L''.)
We will reformulate the theory in terms of currents (key to
bosonization).  Consider: \begin{eqnarray}
&&J(x)~~J(x+\epsilon)~~~~~\mbox{as}~~~~\epsilon\rightarrow
0\nonumber\\  =&& :\psi^\dagger (x)\psi (x)\psi^\dagger (x+\epsilon
)\psi (x+\epsilon ):\nonumber \\ +&&[:  \psi^\dagger
(x)\psi(x+\epsilon):+:\psi (x)\psi^\dagger
(x+\epsilon):]G(\epsilon)+ G(\epsilon)^2\nonumber\\ G(\epsilon)= &&
\langle 0|\psi(x)\psi^\dagger (x+\epsilon)|0\rangle
=\frac{1}{-i\epsilon} .\end{eqnarray} By Fermi statistics the
4-Fermi term vanishes as $\epsilon\rightarrow 0$ \begin{equation}
:\psi^\dagger (x)\psi(x)\psi^\dagger (x)\psi(x):\  =-:\psi^\dagger
(x)\psi^\dagger (x)\psi(x)\psi(x):\ =0 .\end{equation} The second
term becomes a derivative,  \begin{eqnarray} \lim_{\epsilon
\rightarrow  0}[J(x)J(x+\epsilon )+\frac{1}{\epsilon^2}]
&=&\lim_{\epsilon\rightarrow  0}\frac{1}{-i\epsilon}[:\psi^\dagger
(x)\psi(x+\epsilon):- :\psi^\dagger
(x+\epsilon)~~\psi(x):]\nonumber\\ &=&2i:\psi^\dagger
\frac{d}{dx}\psi:\nonumber\\ {\cal H}&=&\frac{1}{4\pi}
J(x)^2+\mbox{constant} .\end{eqnarray} Now consider the commutator,
$[J(x),J(y)]$. The quartic and quadratic terms cancel. We must be
careful about the divergent  c-number part,  \begin{eqnarray}
[J(x),J(y)]&=&
-\frac{1}{(x-y-i\delta)^2}+\frac{1}{(x-y+i\delta)^2}\ \
(\delta\rightarrow  0^+)\nonumber\\ &=&\frac{d}{dx}\left[
\frac{1}{x-y-i\delta}-\frac{1}{x-y+i\delta}\right] \nonumber\\
&=&2\pi i \frac{d}{dx} \delta(x-y) \end{eqnarray}

Now consider the free massless boson theory with Hamiltonian
density (setting $v_F=1$): \begin{equation} {\cal
H}=\frac{1}{2}\left( \frac{\partial\phi}{\partial
t}\right)^2+\frac{1}{2}\left( \frac{\partial\phi}{\partial
x}\right)^2,
 ~~~~[\phi(x), \frac{\partial}{\partial t}\phi (y)]=i\delta(x-y)
\end{equation}

We can again decompose it into the left and  right-moving parts,
\begin{eqnarray} ({\partial_t}^2-{\partial_x}^2)\phi&=&
(\partial_t+\partial_x)(\partial_t-\partial_x)\phi \nonumber\\
\phi(x,t)&=&\phi_L(x+t)+\phi_R(x-t)\nonumber\\
(\partial_t-\partial_x)\phi_L&\equiv&\partial_-\phi_L=0,
{}~~\partial_+\phi_R=0\nonumber\\
H&=&\frac{1}{4}(\partial_-\phi)^2+\frac{1}{4}(\partial_+\phi)^2=
\frac{1}{4}(\partial_-\phi_R)^2+\frac{1}{4}(\partial_+\phi_L)^2
\end{eqnarray}

Consider the Hamiltonian density for a left-moving boson field:
\begin{eqnarray} {\cal H}&=&\frac{1}{4}(\partial_+\phi_L)^2
\nonumber\\ {[}\partial_+\phi_L(x), \partial_+\phi_L(y)]&=&
[\dot\phi+\phi',\dot\phi+\phi']=2i\frac{d}{dx}\delta(x-y)
\end{eqnarray}

Comparing to the Fermionic case, we see that: \begin{equation}
J_L=\sqrt{\pi}\partial_+\phi_L=\sqrt{\pi}\partial_+\phi,
\end{equation} since the commutation relations and Hamiltonian are
the  same. That means the operators are the same  with  appropriate
boundary conditions.

\begin{figure}
\epsfxsize=10 cm
\centerline{\epsffile{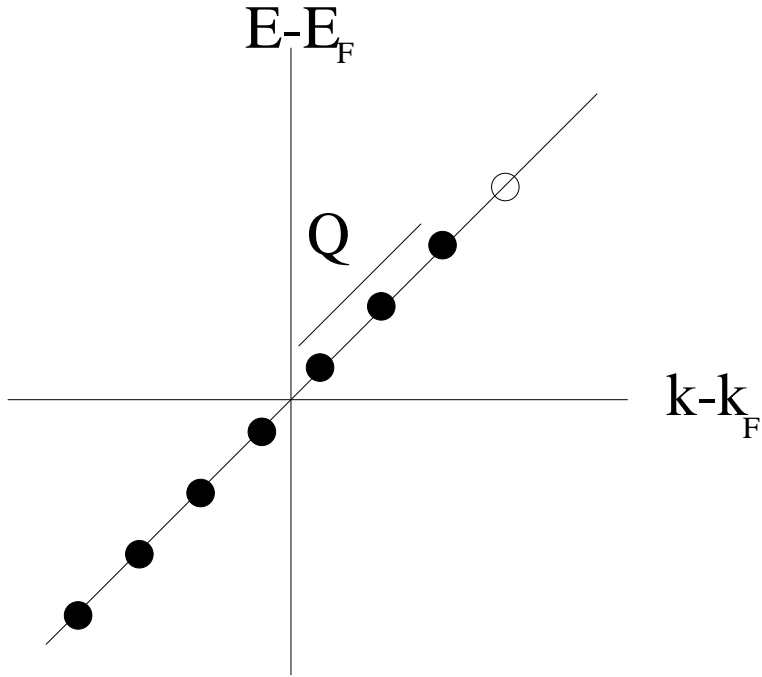}}
\caption{A state with $Q$ extra electrons added to the
groundstate.} \label{fig:Q}
\end{figure}

\begin{figure}
\epsfxsize=10 cm
\centerline{\epsffile{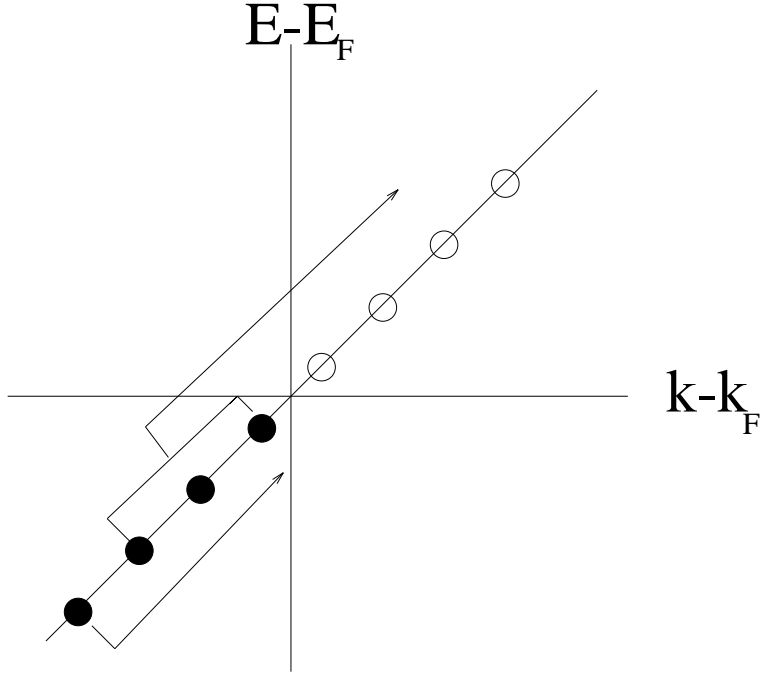}}
\caption{A particle-hole excitation in which three
electrons are raised  four levels and then one
electron is raised three levels.}
\label{fig:ph}
\end{figure}

 Let's compare the spectra. For the Fermionic case, choose boundary
condition: \begin{equation}
\psi(l)=-\psi(-l)~~~~(i.e.~~\psi_L(l)+\psi_R(l)=0),~~~~
k=\frac{\pi}{l}(n+\frac{1}{2}),~~ n=0,\pm 1,\pm 2...\end{equation}
[See Figure (\ref{fig:anti}).  Note that we have shifted $k$ by
$k_F$.]  Consider the minimum energy state of charge $Q$ (relative
to the ground  state). See Figure (\ref{fig:Q}).   We have the
single Fermion energy:  \begin{equation} E=v_Fk,\end{equation}
so:\begin{equation}
E(Q)=v_F\frac{\pi}{l}\sum^{Q-1}_{n=0}(n+\frac{1}{2})=\frac{v_F
\pi}{2l}Q^2 .\label{E(Q)}\end{equation}   Now consider particle
hole excitations relative to the $Q$-ground state:
 The most general particle-hole excitation is obtained by raising
$n_m$ electrons by $m$ levels, then $n_{m-1}$ electrons by $m-1$
levels, etc. [See Figure (\ref{fig:ph}).] \begin{equation}
E=\frac{\pi v_F}{l}(\frac{1}{2}Q^2+\sum^\infty_{m=1}n_m\cdot m)
\label{1spectrum}\end{equation}

Now consider the bosonic spectrum. What are the  boundary
conditions?  Try the periodic one,  \begin{equation}
\phi(l)=\phi(-l)\Rightarrow k=\frac{\pi m}{l} \end{equation} The
$m^{th}$ single particle level has $E_m=v_Fk_m$. The total energy
is \begin{equation} E=\frac{\pi v_F}{l}(\sum^{\infty}_1 n_m \cdot
m),~~~~ n_m= \mbox{occupation~~number}:~0,1,2,... \end{equation}
Where does the $Q^2$ term in Eq. (\ref{1spectrum}) come from?  We
need more general boundary condition on the boson field. Let $\phi$
be an angular variable: \begin{eqnarray}
&&\phi_L(-l)=\phi_L(l)+\sqrt{\pi}Q,~~~~~ Q=0,\pm
1,\pm2,...\nonumber\\ \Rightarrow
&&\phi_L(x+t)=\frac{\sqrt{\pi}}{2}\frac{Q}{l}\cdot(x+t)+\sum^\infty_{m=1}
{_\frac{1}{\sqrt{4\pi  m}} (e^{-i\frac{\pi m}{l}(x+t)} a_m+h.c.)}
,\end{eqnarray} where $a_n$'s are the annihilation operators and
$Q$ is the winding number, \begin{equation} E=\int^l_{-l}
dx[\frac{1}{2}\left( \frac{\partial\phi}{\partial
t}\right)^2+\frac{1}{2}\left( \frac{\partial\phi}{\partial
x}\right)^2]=\frac{\pi}{l}[\frac{1}{2}Q^2+...]. \end{equation} Here
we have set $v_F=1$. We have the following correspondence:\\
soliton $\leftrightarrow$ electron,~~~ oscillator $\leftrightarrow$
particle-hole pair.\\ It is also possible to represent fermion
operators in terms of the boson, \begin{equation} \psi_L\sim
e^{i\sqrt{4\pi}\phi_L}, \end{equation} which gives the correct
Green's function and implies the same angular definition of
$\phi_L$.

 For the Kondo effect we are also interested in the phase-shifted
boundary condition: [See Figure (\ref{fig:per}).] \begin{eqnarray}
\psi_L(l)&=& + \psi_L(-l),~~~~~~~~~k=\frac{\pi}{l}n,~~~ \mbox{(for
fermions)} \nonumber\\ E&=&\frac{\pi v_F}{l}\left[
\frac{Q(Q-1)}{2}+\sum^\infty_1 n_m m\right] . \end{eqnarray} We
have the degenerate ground state, $Q=0$ or $1$, which correspond to
an anti-periodic boundary condition on $\phi$, \begin{eqnarray}
\phi(l)&=&\phi(-l)+\sqrt{\pi}(Q-\frac{1}{2})\nonumber\\
E&=&\frac{\pi}{l}\frac{1}{2}(Q-\frac{1}{2})^2+...=\frac{\pi}{l}
(\frac{1}{2}Q(Q-1)+\mbox{const.}  +...) \end{eqnarray}

Now we  include spin, i.e. we have $2$-component electrons,
\begin{equation} H_0=iv_F\psi^{\alpha
\dagger}\frac{d}{dx}\psi_\alpha,~~~~~~~(\alpha=1,2,~~~\mbox{summed}).
\end{equation} Now we have charge and spin currents (or densities).
We can write H in a manifestly $SU(2)$ invariant way, quadratic in
charge and spin currents: \begin{equation}
J=:\psi^{\alpha\dagger}\psi_{\alpha}:~~,~~~~\vec{J}=\psi^{\dagger\alpha}
\frac{\vec{\sigma}_\alpha^\beta}{2}\psi_{\beta} \end{equation}
Using: \begin{eqnarray}
\vec{\sigma}_\alpha^\beta\cdot\vec{\sigma}_\gamma^{\delta}
&=&2\delta^{\beta}_{\gamma}\delta^{\delta}_{\alpha}-
\delta^\alpha_\beta\delta^\gamma_\delta\\
{\vec{J}}^2&=&-\frac{3}{4}:\psi^{\dagger\alpha}\psi_{\alpha}
\psi^{\dagger\beta}\psi_{\beta}:+\frac{3i}{2}\psi^{\alpha
+}\frac{d}{dx}\psi_{\alpha}+c\mbox{-number},\nonumber\\
J^2&=&:\psi^{\dagger
\alpha}\psi_{\alpha}\psi^{\dagger\beta}\psi_{\beta}:+2i\psi^{\alpha
+}\frac{d}{dx}\psi_{\alpha}+ c\mbox{-number},\nonumber\\ {\cal
H}&=&\frac{1}{8\pi}J^{2}+\frac{1}{6\pi}{\vec{J}}^2, \end{eqnarray}
we have the following commutation relations, \begin{eqnarray}
[J(x),J(y)]&=&4\pi i\delta '(x-y), \mbox{(twice the result for the
spinless case)} \nonumber\\
{[}J(x),J^z(y)]&=&\frac{1}{2}[J_{\uparrow}+J_{\downarrow},J_{\uparrow}-
J_{\downarrow}]=0  .\end{eqnarray} From $[J,\vec{J}]=0$, we see
that  $H$ is sum of commuting charge and spin parts.
\begin{eqnarray} [J^a(x),J^b(y)]&=&2\pi\psi^\dagger
[{\frac{\sigma}{2}}^a,{\frac{\sigma}{2}}^b]
{\psi}\cdot\delta(x-y)+tr[{\frac{\sigma}{2}}^a,{\frac{\sigma}{2}}^b]2\pi
i \frac{d}{dx}\delta(x-y)\nonumber\\  &=&2\pi i
\epsilon^{abc}J^c(x)\cdot\delta(x-y)+\pi i\delta^{ab}
\frac{d}{dx}\delta(x-y) .\end{eqnarray} We obtain the Kac-Moody
algebra of central charge $k=1$. More generally the  coefficient of
the second term is multiplied by an integer $k$. \\ Fourier
transforming, \begin{equation} \vec J_n
\equiv\frac{1}{2\pi}\int^l_{-l} dx e
^{in\frac{\pi}{l}x}\vec{J}(x),~~~~
[J^a_n,J^b_m]=i\epsilon^{abc}J^c_{n+m}+\frac{1}{2}n\delta^{ab}\delta_{n,-m}
\end{equation} we have an $\infty$-dimensional generalization of
the ordinary $SU(2)$ Lie algebra. The spin part of the  Hamiltonian
is  \begin{equation}
H_{s}=\frac{\pi}{l}\frac{1}{3}\sum^{\infty}_{n=-\infty}:\vec J_{-n}
\cdot\vec J_{n}: \end{equation} The spectrum of $H_{s}$ is again
determined by the algebra obeyed by the $\vec{J}_{n}$'s  together
with boundary conditions. The construction is similar to building
representations of $SU(2)$ from  commutation relations, i.e.
constructing raising operator, etc.

In the $k=1$ case we are considering here it is simplest to use:
\begin{eqnarray} \vec{J}(x)^2 &=&3(J^z(x))^2\nonumber\\
 {\cal H} &=&\frac{1}{8\pi}J^2+\frac{1}{2\pi}(J^z)^2\nonumber\\
   &=&\frac{1}{4\pi}(J^2_{\uparrow}+J^2_{\downarrow})\nonumber\\
&=&\frac{1}{4}((\partial_+\phi_\uparrow)^2+(\partial_+\phi_\downarrow)^2)
\nonumber\\
&=&\frac{1}{4}[(\partial_+(\frac{\phi_\uparrow+\phi_\downarrow}{\sqrt{2}}))^2+
(\partial_+(\frac{\phi_\uparrow-\phi_\downarrow}{\sqrt{2}}))^2]\nonumber\\
   &=&\frac{1}{4}({(\partial_+\phi_c)}^2+{(\partial_+\phi_s)}^2)
\end{eqnarray} Now we have introduced two commuting charge and spin
free massless bosons. SU(2) symmetry is now concealed but boundary
condition on $\phi_s$ must  respect it. Consider the spectrum of
fermion theory with boundary  condition: $\psi(l)=-\psi(-l)$,
\begin{equation} E=\frac{\pi  V}{l}\left[
{\frac{Q_\uparrow}{2}}^2+{\frac{Q_\downarrow}{2}}^2+
\sum^\infty_{m=-\infty}m(n^\uparrow_m+n^\downarrow_m)\right] .
\end{equation}  Change over to $\phi_c$ and $\phi_s$. We can
relabel occupation numbers,  \begin{eqnarray}
n^\uparrow_m,~n^\downarrow_m~&\longrightarrow&~n^c_m,~n^s_m\nonumber\\
Q&=&Q_\uparrow +Q_\downarrow\nonumber\\
S^z&=&\frac{1}{2}(Q_\uparrow-Q_\downarrow)\nonumber\\ E&=&\frac{\pi
v_F}{l}[\frac{1}{4}Q^2+{(S^z)}^2+\sum^\infty_1
mn^c_m+\sum^\infty_1mn_m^s]\label{EcEs}\\ &=&E_c+E_s\nonumber\\
\phi_c&=&\frac{\sqrt{\pi}}{2\sqrt{2}}\frac{Q}{l}(x+t)+...\nonumber\\
\phi_s&=&\frac{\pi}{\sqrt{2}}{\frac{S^z}{l}}(x+t)+...
\end{eqnarray} Actually charge and spin bosons are not completely
decoupled; we must require $Q=2S^z$ (mod $2$), to correctly
reproduce the free fermion spectrum.  We see that the boundary
conditions on $\phi_c$ and $\phi_c$ are coupled. Now consider the
phase-shifted case. \begin{equation} E=\frac{\pi
v_F}{l}[\frac{1}{4}(Q-1)^2+(S^z)^2+...] \end{equation} Redefine
$Q-1\rightarrow Q$ so  \begin{equation} E=\frac{\pi
v_F}{l}[\frac{1}{4}Q^2+(S^z)^2+....] ,\end{equation} the same as
before the phase shift, Eq. (\ref{EcEs}).  One of the $0$-energy
single-particle states is filled, for $Q=0$ and there are 4
groundstates, \begin{equation} (Q,S^z)=(0,\pm \frac{1}{2}),~~(\pm
1,0) .\end{equation} Now  $Q=2S^z+1$ (mod $2$); i.e. we ``glue"
together charge and spin excitations in two different  ways,
either  \begin{eqnarray} \hbox{(even, integer)}&\oplus& \hbox{(odd,
half-integer)}\nonumber \\
 \hbox{or\ \  (even, half-integer)} &\oplus& \hbox{(odd,
integer)},\end{eqnarray} depending on the boundary conditions.  The
$\frac{\pi}{2}$ phase shift simply reverses these ``gluing
conditions''.

The set of all integer spin states form a ``conformal tower". They
can be  constructed from the Kac-Moody algebra by  applying the
raising operators $\vec J_{-n}$ to the lowest (singlet) state, with
all spacings  $\frac{\pi  v_F}{l}\cdot$(integer). Likewise for all
half-integer spin states,  $(s^z)^2=\frac{1}{4}+$integer. Likewise
for even and odd charge states.  The K-M algebra determines
uniquely conformal towers but  boundary conditions  determine which
conformal towers occur in the spectrum and in which spin-charge
combinations.  \section{Conformal Field Theory Approach to The
Kondo Effect} The chiral one-dimensional Hamiltonian density of Eq.
(\ref{Hochi}) and (\ref{HINT}) is: \begin{equation} {\cal
H}=\frac{d}{dx}\psi_{L\alpha}+
\lambda\psi^{\dagger\alpha}_L\frac{\vec{\sigma}_\alpha^\beta}{2}
\psi_{L\beta}\cdot\vec{S}~\delta(x)~~~ \mbox{(left-movers only)}
\end{equation} We rewrite it in terms of spin and charge currents
only, \begin{equation} {\cal
H}=\frac{1}{8\pi}J^2+\frac{1}{6\pi}(\vec{J})^2+\lambda\vec{J}\cdot\vec{S}
{}~\delta(x) .\end{equation} The Kondo interaction involves spin
fields only, not charge fields: $H=H_s+H_c.$ Henceforth we only
consider the spin part. In Fourier transformed form,
\begin{eqnarray}
H_s&=&\frac{\pi}{l}(\frac{1}{3}\sum^\infty_{n=-\infty}\vec
J_{-n}\cdot \vec J_n+\lambda\sum^\infty_{n=-\infty}\vec
J_n\cdot\vec{S})\nonumber\\ {[}J^a_n,
J^b_m]&=&i\epsilon^{abc}J^{c}_{n+m}+\frac{n}{2}\delta^{ab}\delta_{n,-m}
\nonumber\\ {[} S^{a}, S^{b} ]
 &=&i \epsilon^{abc}S^c \nonumber\\ {[} S^a , J^{b}_{n} ]  &=&0
\end{eqnarray}
 {}From calculating Green's functions for $\vec{J}(x)$ we could
again reproduce  perturbation theory $\frac{d\lambda}{d
lnD}=-\lambda^2+\cdots $. That is a small  $\lambda >0$ grows. What
is the infrared stable fixed point? Consider $\lambda=\frac{2}{3}$,
where we may ``complete the square".  \begin{eqnarray}
H&=&\frac{\pi  V}{3l}\sum^{\infty}_{n=-\infty}[(\vec J_{-n}+\vec{S})
\cdot(\vec J_n+\vec{S})-\frac{3}{4}]\nonumber\\ {[}J^a_n+S^a,
J^b_m+S^b]&=&i\epsilon^{abc}(J^c_{n+m}+S^c)+
\frac{n}{2}\delta^{ab}\delta_{n,-m} .\end{eqnarray} $H$ is
quadratic in the new currents, $\vec{{\cal J}}_n\equiv\vec
J_n+\vec{S}$, which  obey the same Kac-Moody algebra! What is the
spectrum of  $H(\lambda=\frac{2}{3})?$ We must get back to
Kac-Moody conformal towers  for integer  and half-integer spin.
This follows from the KM algebra and the form of $H$ (i.e. starting
{}from the lowest state we  produce the entire tower by applying the
raising operators, $\vec{J}_{-n}$).

\begin{figure}
\epsfxsize=10 cm
\centerline{\epsffile{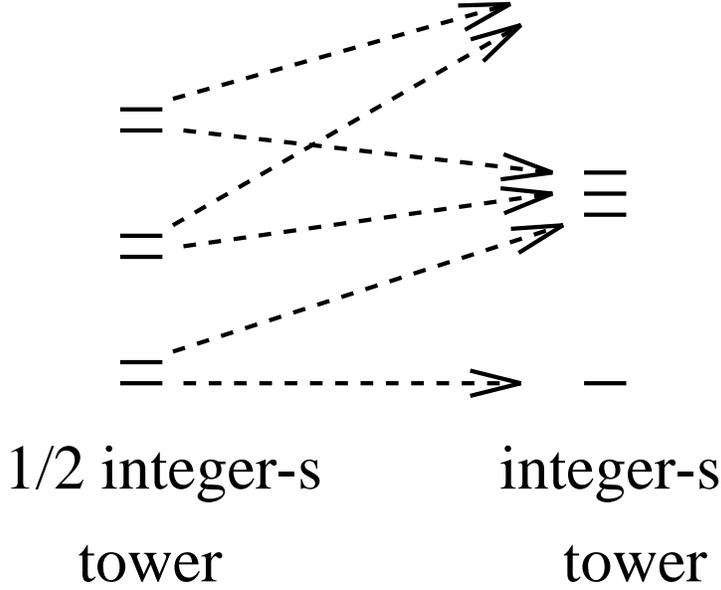}}
\caption{At $\lambda = 2/3$ the 1/2-integer-spin conformal tower is
mapped into the integer-spin conformal tower.} \label{fig:square}
\end{figure}

Thus we find that {\it the strong-coupling fixed point is the same
as the  weak-coupling fixed point}. However, the total spin
operator is now $\vec{\cal J}_0=\vec J_0+\vec{S}$.  We consider
impurity spin magnitude, s=1/2. Any integer-spin state becomes a
1/2-integer spin state and vice versa. \begin{equation}
\hbox{Integer}\leftrightarrow \hbox{1/2-Integer}.\end{equation}
 Presumably $\lambda=2/3$ is the strong coupling fixed point in
this  formulation of the problem. $\infty$-coupling can become
finite coupling under  a redefinition, eg. \begin{equation}
\lambda_{\hbox{Lattice}}=\frac{\lambda_{KM}}{1-\frac{3}{2}\lambda_{KM}}.
\end{equation} We  expect the low-energy, large $l$ spectrum to be
KM conformal towers for any  $\lambda$. The effect of the Kondo
interaction is to interchange the two conformal towers, Integer
$\Leftrightarrow ~~ \frac{1}{2}$-Integer. See Figure
(\ref{fig:square}). This is equivalent to a $\frac{\pi}{2}$
phase-shift,  \begin{eqnarray} \hbox{(even,
integer)}&\oplus&\hbox{(odd,$\frac{1}{2}$-integer)}  \nonumber \\
&\Updownarrow&\nonumber \\ \hbox{(even,
$\frac{1}{2}$-integer)}&\oplus&\hbox{(odd,integer)}\end{eqnarray}
\subsection{Leading Irrelevant Operator:Specific Heat,
Susceptibility,
 Wilson Ratio, Resistivity at $T>0$}
 At the stable fixed point $\vec{S}$ has  disappeared; i.e. it is
absorbed into $\vec{\cal J}$, \begin{equation} \vec{\cal
J}(x)=\vec{J}(x)+2\pi\vec{S}~\delta(x). \end{equation} What
interactions could be generated in $H_{eff}$?  These only involve
$\vec{\cal J}$, not $\vec{S}$.  \begin{equation}
H_s=\frac{1}{6\pi}\vec{\cal J}(x)^2+\lambda_1\vec{\cal
J}(0)^2\delta(x). \end{equation} This is the only dimension-2
rotationally invariant operator  in the spin sector. We  have
succeeded in reducing two dimension-2 operators to one. The  other
one is the charge-operator $\lambda_2J(0)^2\delta(x)$, $\lambda_2=0$
because there is no interaction in the charge sector (with other
regularization we expect $\lambda_1\sim\frac{1}{T_K}$,
$\lambda_2\sim\frac{1}{D}<<\lambda_1$).

 Now we calculate the specific heat and  susceptibility to $1st$
order  in $ \lambda_1$.\\ Susceptibility of left-moving free
fermions: \begin{eqnarray} 0\mbox{-th order}~~~~
M&=&\frac{1}{2}(n_{\uparrow}-n_{\downarrow}) =l\int d\epsilon\
\nu(\epsilon)[n(\epsilon+\frac{h}{2})-n(\epsilon-\frac{h}{2})]\nonumber\\
\chi&=&\frac{l}{2\pi}~~~~(for~~~~ T<<D)\nonumber\\ 1\mbox{st
order}~~~~ \chi&=&\frac{1}{3T}\langle[\int dx\vec{\cal
J}(x)]^2\rangle_{\lambda_1} \nonumber\\
    &=&\chi_0-\frac{\lambda_1}{3T^2}\langle[\int dx  \vec{\cal
J}(x)]^2\vec{\cal J}(0)^2\rangle+... \end{eqnarray} A simplifying
trick is to replace: \begin{equation} \delta  {\cal
H}=\lambda_1\vec{\cal J}^2(0)\delta(x)\longrightarrow
\frac{\lambda_1}{2l}\vec{\cal J}^2(x), \end{equation} which gives
the same result to first order in $\lambda$ (only) by
translational  invariance of $H$ at $\lambda=0$. Now the
Hamiltonian density changes into \begin{equation} {\cal
H}\rightarrow(\frac{1}{6\pi}+\frac{\lambda_1}{2l})~~\vec{\cal
J}^2(x) .\end{equation} We simply rescale $H$ by a factor
\begin{equation} H\rightarrow(1+\frac{3\pi\lambda_1}{l})H
.\end{equation} Equivalently in a thermal average,
\begin{equation} T\rightarrow\frac{T}{1+\frac{3\pi\lambda_1}{l}}
\equiv T(\lambda_1) \end{equation} \begin{eqnarray}
\chi(\lambda_1,T)&=&\frac{1}{3T}<(\int\vec{\cal
J})^2>_{T(\lambda_1)} \nonumber\\
&=&\frac{1}{1+3\pi\lambda_1/l}\chi(0,T(\lambda_1))\nonumber\\
&\approx& [1-\frac{3\pi\lambda_1}{l}]\chi_0\nonumber\\
&=&\frac{l}{2\pi}-{3\lambda_1\over 2}, \end{eqnarray} where in the
last equality the first term represents the bulk part  and the
second one, of order $\sim\frac{1}{T_K}$, comes from the impurity
part.    Specific Heat:\\ \begin{equation} 0\mbox{-th order}~~~~
C=C_c+C_s,~~~ C_c=C_s=\frac{\pi l T}{3}. \end{equation} Each free
left-moving boson makes an identical contribution. \begin{eqnarray}
1\mbox{st order in } \lambda_1 ~~~~
C_s(\lambda_1,T)&=&\frac{\partial}{\partial
T}<H(\lambda_1)>_{\lambda_1}\nonumber\\
&=&C_s(0,T(\lambda_1))\nonumber\\ &=&\frac{\pi
l}{3}\frac{T}{1+3\pi\lambda_1/l}\nonumber\\ &\approx&\frac{\pi l
T}{3}-\pi^2\lambda_1 T \end{eqnarray} \begin{equation} \frac{\delta
C_s}{C_s}=-\frac{3\pi\lambda_1}{l}=2\frac{\delta C_s}{C}
\end{equation} The Wilson Ratio: \begin{equation}
R_w\equiv\frac{\delta\chi/\chi}{\delta  C/C}=2=\frac{C}{C_s}
\end{equation} measues the fraction of $C$ coming from the spin
degrees of freedom.

\begin{figure}
\epsfxsize=10 cm
\centerline{\epsffile{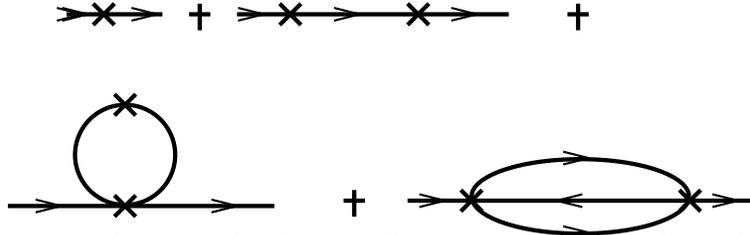}}
\caption{Feynman diagrams contributing to the electron
self-energy up to second order in the leading
irrelevant coupling constant of Eq. (3.8).}
\label{fig:stpert}
\end{figure}

Doing more work, we can calculate the resistivity to
$O(\lambda^2)$.\cite{Nozieres2,Affleck6} First we get  the electron
lifetime from the self-energy. The change in the 3D Green's
function comes  only from the 1D s-wave part: \begin{eqnarray}
&&G_3(\vec{r_1},\vec{r_2})-G^0_3(|\vec{r_1}-\vec{r_2}|)\nonumber \\
=&&\frac{1}{8\pi^2r_1r_2}[e^{-ik_F(r_1+r_2)}(G_{LR}(r_1,r_2)-G_{LR,0}(r_1,r_2))
+h.c.]\nonumber\\ =&&G^0_3(r_1)\Sigma G^0_3(r_2). \end{eqnarray}
The self-energy $\Sigma$ depends only on the frequency. It gets
multiplied by  the impurity concentation for a finite density (in
the dilute limit). We must calculate the 1D Green's function
$G_{LR}(r_1,r_2,\omega) $  perturbatively in $\lambda$
\begin{eqnarray} O(\lambda^0_1):~~~~ G_{LR}(r_1,r_2)
&=&-G^0_{LL}(r_1,-r_2)\nonumber\\ &=&-G^0_{LL}(r_1+r_2)\nonumber\\
&=&-G^0_{LR}(r_1,r_2) ,\end{eqnarray} where the  ($-$) sign comes
{}from the change in boundary conditions, \begin{equation}
G_{LR}-G^0_{LR}=-2 G^0_{LR}+O(\lambda_1) \end{equation}  To
calculate to higher orders it is convenient to write the
interaction as: \begin{equation} ~~ {\vec{\cal
J}}^2=-\frac{3}{4}:\psi^{\dagger\alpha}
\psi_{\alpha}\psi^{\dagger\beta}\psi_{\beta}:
+\frac{3i}{4}(\psi^{\dagger\alpha}\frac{d}{dx}\psi_{\alpha}
-\frac{d\psi^{\dagger \alpha}}{dx}\psi_{\alpha}) \end{equation} To
second order in $\lambda_1$, we have the Feynman diagrams shown in
Figure (\ref{fig:stpert}), giving: \begin{equation}
\Sigma^R(\omega)=\frac{-in}{2\pi\nu}[2+3\pi
i\lambda_1\omega-\frac{1}{2}(3\pi\lambda_1)^2\omega^2
-\frac{1}{4}(3\pi\lambda_1)^2(\omega^2+\pi^2T^2)] .\end{equation}
The first three terms give a phase-shift and the last term
represents inelastic scattering.   \begin{eqnarray}
\Sigma^R(\omega)&=&\frac{-in_i}{2\pi\nu}[1-e^{2i\delta(\omega)}]+
\Sigma^R_{inel}(\omega)\nonumber\\
\delta&=&\frac{\pi}{2}+\frac{3\pi\lambda_1\omega}{2}+...\nonumber\\
 \frac{1}{\tau}&=&-2 I_m\Sigma_R(\omega)\nonumber\\
\frac{1}{\tau(\omega)}&=&\frac{n_i}{\pi\nu}[2-\frac{1}{2}(3\pi\lambda_1)^2
\omega^2-\frac{1}{4}(3\pi\lambda_1)^2(\omega^2+\pi^2T^2)]
\end{eqnarray} The leading $\lambda _1$ dependence is
$O(\lambda^2_1)$ in this case.  The $O(\lambda_1)$ term in
$\Sigma^R$ is real. We calculate the conductivity from the Kubo
formula. (There is no contribution from the scattering vertex for
pure s-wave scattering.) \begin{eqnarray} \sigma(T)&=&\frac{2
e^2}{3m^3}\int\frac{d^3\vec{k}}{(2\pi)^3}\left[-\frac{\partial
n}{\partial\epsilon_k}\right]{\vec{k}}^2\tau(\epsilon_k)\nonumber\\
\tau(\epsilon_k)&\approx&\frac{\pi\nu}{2n_i}[1+\frac{1}{4}(3\pi{\lambda_1}^2)
\epsilon_k^2+\frac{1}{8}(3\pi\lambda_1)^2(\epsilon_k^2+(\pi^2T^2)]
\nonumber\\ \rho(T)&
=&\frac{1}{\sigma(T)}=\frac{3n_i}{\pi(ev_F\nu)^2}[1-\frac{9}{4}\pi^4
\lambda_1^2T^2] \end{eqnarray} All low-temperature properties are
determined in terms of one unknown coupling  constant
$\lambda_1\sim\frac{1}{T_K}$. Numerical or Bethe ansatz methods are
needed to  find the precise value of
$\lambda_1(D,\lambda)\propto\frac{1}{D}e^{1/\lambda}$.

\section{Multi-Channel Kondo Effect}  Normally there are several
``channels"  of electrons -e.g. different d-shell orbitals. A very
simple and symmetric model is: \begin{equation} H=\sum_{\vec
k,\alpha ,  i=1,2,...k}\epsilon_{\vec k}\psi^{\dagger\alpha
i}_{\vec k} \psi_{\vec k \alpha
 i}+\lambda\vec{S}\cdot\sum_{\vec k,\vec k'\\ \alpha,\beta
i}\psi^{\dagger\alpha  i}_{\vec
k}\vec{\sigma}_\alpha^{\beta}\psi_{\vec k' \beta i}. \end{equation}
This model has $SU(2)\times SU(k)\times U(1)$ symmetry.  Realistic
systems do not have this full symmetry.  To understand the
potential applicability of this model we need to analyse the
relevance of various types of symmetry breaking.\cite{Affleck5}  An
interesting possible experimental application of the model was
proposed by Ralph, Ludwig, von Delft and Buhrman.\cite{Ralph1}  In
general, we let the impurity have an arbitrary spin, s, as well.

\begin{figure}
\epsfxsize=10 cm
\centerline{\epsffile{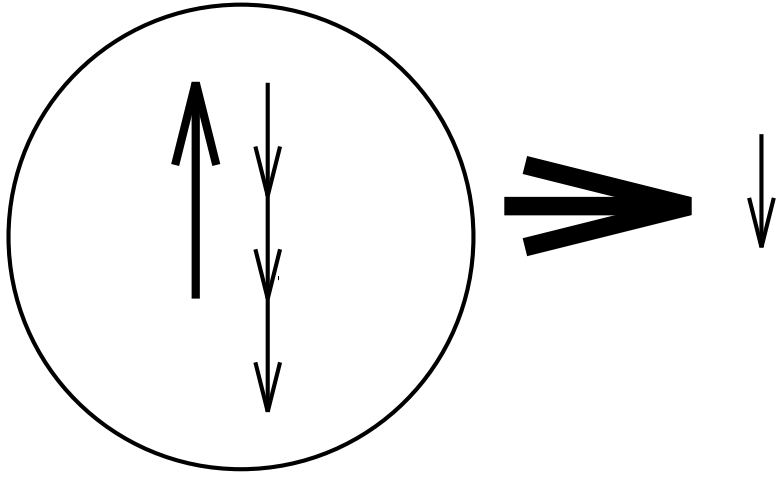}}
\caption{Formation of an effective spin at strong
Kondo coupling.  $k=3$, $s=1$ and
$s_{\hbox{eff}}=1/2$.}
\label{fig:seff}
\end{figure}

 Perturbation theory in $\lambda$ is similar to the result
mentioned before: \begin{eqnarray} \frac{d\lambda}{d~ln
D}&=&-\nu\lambda^2+\frac{k}{2}\nu^2\lambda^3+O[ks(s+1)\lambda^4]\nonumber\\
{\vec{S}}^2&=&s(s+1) .\end{eqnarray} Does
$\lambda\rightarrow\infty$ as $T\rightarrow 0$? Let's suppose it
does and  check consistency. What is the  groundstate for the
lattice model of Eq. (\ref{lattice}), generalized to arbitrary $k$
and $s$, at $\lambda/t \rightarrow\infty$?  In the limit we just
consider the single-site model:  \begin{equation}
H=\lambda\vec{S}\cdot\psi^\dagger_0\frac{\vec{\sigma}}{2}\psi_0
,\end{equation} for $\lambda>0$ (antiferromagnetic case) the
minimum energy state has maximum spin for electrons at  $\vec{0}$
i.e. spin$=k/2$.  Coupling this spin-k/2 to a spin-s, we don't get
a singlet  if $s\neq k/2$, but rather an effective spin of size
$|s-k/2|$. [See Figure (\ref{fig:seff}).] The impurity is
underscreened $(k/2<s)$  or overscreened $(k/2>s)$.

\begin{figure}
\epsfxsize=10 cm
\centerline{\epsffile{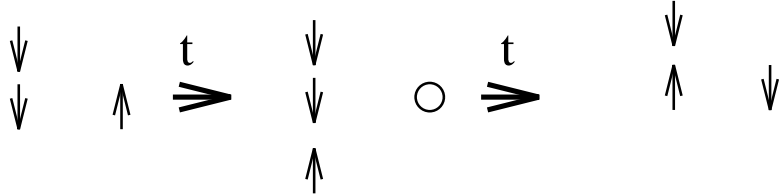}}
\caption{Effective Kondo interaction of $O(t^2)$.}
\label{fig:exeff}
\end{figure}

\begin{figure}
\epsfxsize=10 cm
\centerline{\epsffile{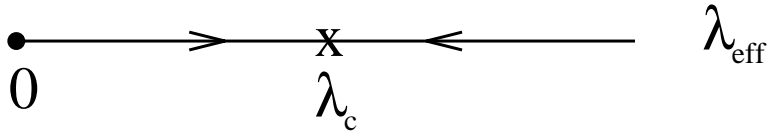}}
\caption{RG flow of the Kondo coupling in the overscreened case.}
\label{fig:flowov}
\end{figure}

Now let $\frac{t}{\lambda}<<1 $ be finite. Electrons on site $\pm
1$ can  exchange an electron with $0$. This gives an effective
Kondo interaction:
$$\lambda_{\hbox{eff}}\sim{\frac{t}{\lambda}}^2<<1$$  See Figure
(\ref{fig:exeff}). What is the sign of $\lambda_{\hbox{eff}}$? The
coupling of the electron spins is antiferromagnetic:
$\lambda_{\hbox{eff}}\vec S_{e1,0}\cdot \vec S_{e1,1}$, with
$\lambda_{\hbox{eff}}>0$ (as in the  Hubbard model). But we must
combine spins  \begin{equation} \vec S_{\hbox{eff}}=\vec{S}+\vec
S_{el,0}. \end{equation} For $\frac{k}{2}<s, \vec{S}_{eff}\
||-\vec{S}_{el,0}$ but, for $\frac{k}{2}>s, \vec{S}_{eff} \
||+\vec{S}_{el,0}$. So, ultimately, $\lambda_{\hbox{eff}}<0$  in
the underscreened case and $\lambda_{\hbox{eff}}>0$ in the
overscreened case.   In the first (underscreened) case, the
assumption $\lambda\rightarrow\infty$ was consistent since a
ferromagnetic $\lambda_{\hbox{eff}}\rightarrow 0$ under
renormalizaton and this implies $\lambda\rightarrow\infty$, since
$\lambda_{\hbox{eff}}\sim-{\frac{t}{\lambda}}^2$.  In this case we
expect the infrared fixed point to correspond to a decoupled spin
of size $s_{\hbox{eff}}=s-k/2$ and free electrons with a $\pi /2$
phase shift.
 In  the second (overscreened) case the $\infty$-coupling fixed
point is not consistent.  Hence the fixed point occurs at
intermediate  coupling: This fixed point does not correspond to a
simple boundary condition on  electrons, instead it is a
Non-Fermi-Liquid Fixed Point.  See Figure (\ref{fig:flowov}).

\begin{figure}
\epsfxsize=10 cm
\centerline{\epsffile{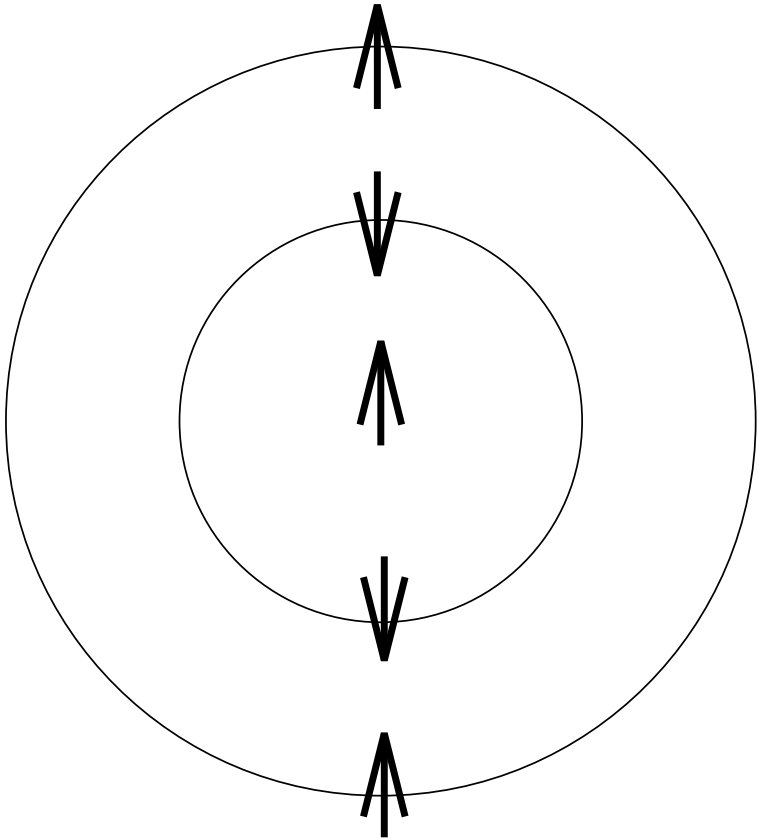}}
\caption{The overscreened case with $s=1/2$, $k=2$.}
\label{fig:s12}
\end{figure}

For the k=2, s=1/2 case we may think of the electrons (one from each
channel) in the first layer around the impurity as aligning
antiferromagnetically with the impurity.  This overscreens it,
leaving an effective s=1/2 impurity.  The electrons in the next
layer then overscreen this effective impurity, etc.  At each stage
we have an effective s=1/2 impurity.  [See Figure
(\ref{fig:s12}).]  Note in this special case that there is a
duality between the weak and strong coupling unstable fixed points:
they both contain an s=1/2 impurity.

\subsection{Large-k Limit} The $\beta$-function is:
\begin{equation} \beta = \lambda^2-{k\over 2}\lambda^3+
O(\lambda^4).\end{equation} If we only consider the first two
terms, there is a fixed point at: \begin{equation} \lambda_c
\approx 2/k. \end{equation} At this (small) value of $\lambda$ the
quartic term, and all higher terms are $O(1/k^4)$, whereas the
quadratic and cubic terms are $O(1/k^2)$.  Thus we may ignore all
terms but the quadratic and cubic ones, for large k.  The slope of
the $\beta$-function at the critical point is: \begin{equation}
\left. {d\beta \over dk}\right|_{\lambda_c}=2\lambda_c -{3\over
2}\lambda^2_c=-{2\over k}.\end{equation} This implies that the
leading irrelevant coupling constant at  the non-trivial (infrared)
fixed point has dimension 2/k at large k, so that
$(\lambda-\lambda_c)$ scales as $\Lambda^{2/k}$.  Thus the leading
irrelevant operator has dimension (1+2/k).  This is not an
integer!  This implies that this critical point is not a Fermi
liquid.

\subsection{Current Algebra Approach} We can gain some insight into
the nature of the non-trivial critical point using the current
algebra approach discussed in the previous section for the k=1
case.  It is now convenient to use a form of bosonization which
separates spin, charge and {\it flavour} (i.e. channel) degrees of
freedom.  This representation is known as a conformal embedding.
We introduce charge ($J$), spin ($\vec J$) and flavour ($J^A$)
currents.  $A$ runs over the $k^2-1$ generators of $SU(k)$.  The
corresponding elements of the algebra are written $T^A$. These are
traceless Hermitean matrices normalized so that: \begin{equation}
\hbox{tr}T^AT^B = {1\over 2} \delta^{AB},\end{equation} and obeying
the completeness relation: \begin{equation} \sum_A
(T^A)_a^b(T^A)^d_c = {1\over 2}\left[ \delta^b_c\delta^d_a -{1\over
k}\delta^b_a\delta^d_c\right] ,\label{complete}\end{equation} and
the commutation relations: \begin{equation} [T^A,T^B] =
if^{ABC}T^C,\end{equation} where the $f^{ABC}$ are the $SU(k)$
structure constants.
 Thus the currents are: \begin{eqnarray} J &\equiv& :\psi^{\dagger
i \alpha}\psi_{i \alpha}:\nonumber \\ \vec J &\equiv& \psi^{\dagger
i \alpha}{\vec \sigma_\alpha^\beta \over 2}\psi_{i \beta}\nonumber
\\ J^A &\equiv& \psi^{\dagger i \alpha}(T^A)_i^j\psi_{j
\alpha}.\end{eqnarray} (All repeated indices are summed.)  It can
be seen using Eq.  (\ref{complete}) that the free fermion
Hamiltonian can be written in terms of these currents as:
\begin{equation} {\cal H} = {1\over 8\pi k}J^2+{1\over 2\pi
(k+2)}\vec J^2+{1\over 2\pi (k+2)}J^AJ^A.\end{equation}  The
currents $\vec J$ obey the $SU(2)$ Kac-Moody algebra with central
charge $k$ and the currents $J^A$ obey the $SU(k)$ Kac-Moody
algebra with central charge 2: \begin{equation}
[J^A_n,J^B_m]=if^{ABC}J^C_{n+m}+n\delta^{AB}\delta_{n,-m}.\end{equation}
  The three types of currents commute with each of the other two
types, as do the three parts of the Hamiltonian.  Thus we have
succeeded in expressing the Hamiltonian in terms of these three
types of excitations: charge, spin and flavour.  The Virasoro
central charge c (proportional to the specific heat) for a
Hamiltonian quadratic in currents of a general group $G$ at level
$k$ is:\cite{Knizhnik} \begin{equation} c_{G,k} =
{\hbox{Dim}(G)\cdot k\over C_V(G)+k},\end{equation} where Dim ($G$)
is the dimension of the group and $C_V(G)$ is the quadratic Casimir
in the fundamental representation.  For $SU(k)$ this has the value:
\begin{equation} C_V(SU(k)) = k.\end{equation}  Thus the total
value of the central charge, c, is: \begin{equation}
c_{\hbox{TOT}}=1+{3\cdot k\over k+2}+{(k^2-1)\cdot 2\over
k+2}=2k,\end{equation} the correct value for 2k species of free
fermions.   Complicated ``gluing conditions'' must be imposed to
correctly reproduce the free fermion spectra, with various boundary
conditions.  These were  worked out in general by Altshuler, Bauer
and Itzykson.\cite{Altshuler}  The $SU(2)_k$ sector consists of
$k+1$ conformal towers, labelled by the spin of the lowest energy
(``highest weight'') state:
$s=0,1/2,1,...k/2$.\cite{Zamolodchikov1,Gepner}

We may now treat the Kondo interaction much as in the single
channel case.  It only involves the spin sector which now becomes:
\begin{equation}{\cal H}_s = {1\over 2\pi (k+2)}\vec J^2 + \lambda
\vec J\cdot \vec S \delta (x).\end{equation} We see that we can
always ``complete the square'' at a special value of $\lambda$:
\begin{equation} \lambda_c={2\over 2+k},\end{equation} where the
Hamiltonian reduces to its free form after a shift of the current
operators by $\vec S$ which preserves the KM algebra.  We note that
at large $k$ this special value of $\lambda$ reduces to the one
corresponding to the critical point:  $\lambda_c\to 2/k.$

While this observation is tantalizing, it leaves many open
questions.  We might expect that some rearranging of the $(k+1)$
$SU(2)_k$ conformal towers takes place at the critical point but
precisely what is it?  Does it correspond to some sort of boundary
condition?  If so what?  How can we calculate thermodynamic
quantities and Green's functions?  To answer these questions we
need to understand some more technical aspects of CFT in the
presence of boundaries. \section{Boundary Conformal Field Theory}
\begin{figure}
\epsfxsize=10 cm
\centerline{\epsffile{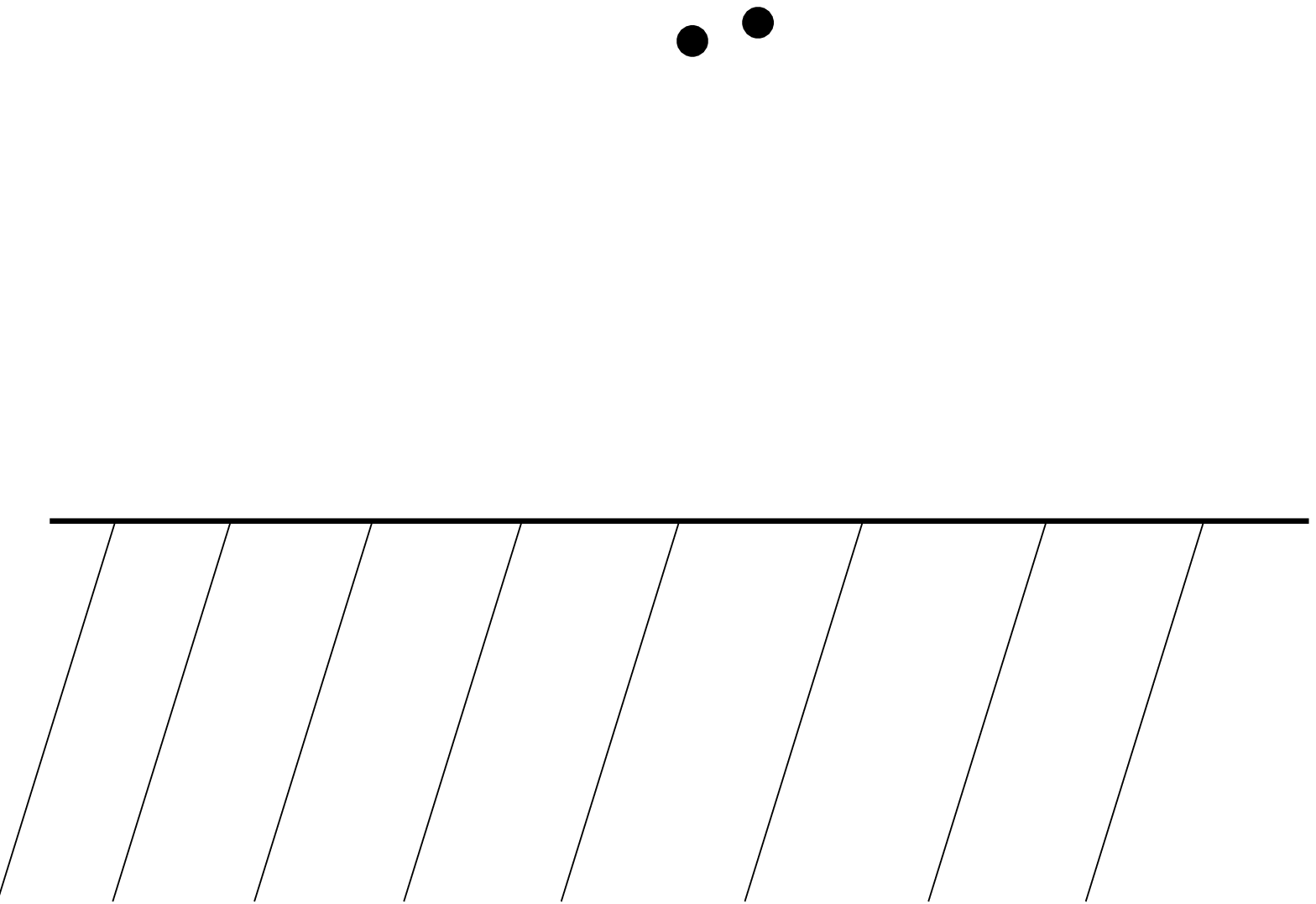}}
\caption{The bulk limit.}
\label{fig:2ptbulk}
\end{figure}

\begin{figure}
\epsfxsize=10 cm
\centerline{\epsffile{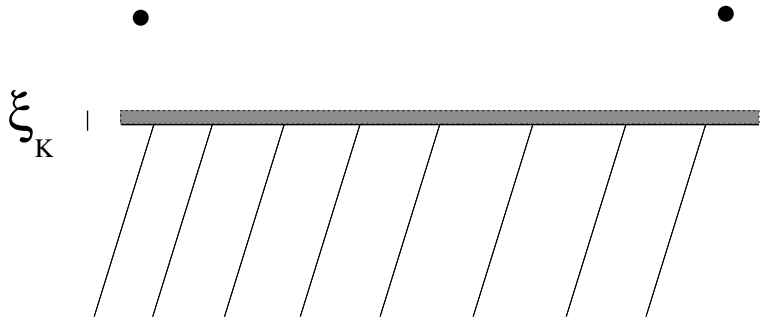}}
\caption{The boundary limit.}
\label{fig:2ptbound}
\end{figure}

We will assume that the critical point corresponds to a conformally
invariant boundary condition on the free theory.  Using the general
theory of conformally invariant boundary conditions developed by
Cardy\cite{Cardy1} we can completely solve for the critical
properties of the model.  Why assume that the critical point
corresponds to such a boundary condition?  It is convenient to work
in the space-(imaginary) time picture.  The impurity then  sits at
the boundary, $r=0$ of the half-plane $r>0$ on which the Kondo
effect is defined. If we consider calculating a two-point Green's
function when both points are taken very far from the boundary
(with their separation held fixed)  then we expect to obtain bulk
behaviour, unaffected by the boundary.  [See Figure
(\ref{fig:2ptbulk}).] This, at long distances and times is the
conformally invariant behaviour of the free fermion system.  Very
close to the boundary, we certainly do not expect the behaviour to
be scale invariant (let alone conformally invariant) because
various microscopic scales become important.  The longest of these
scales is presumably the Kondo scale, $\xi_K\approx v_F/T_L\approx
ae^{1/\nu \lambda}$.  Beyond this distance, it is reasonable to
expect scale-invariant behaviour.  However, if the two points are
far from each other compared to their distance from the boundary
[Figure (\ref{fig:2ptbound})] then the behaviour is still
influenced by the boundary even when both points are far from it.
We have a sort of boundary-dependent termination of the bulk
conformally invariant behaviour.  The dependence on the details of
the boundary (such as the value of $\xi_K$) drops out.  We may
think of various types of boundaries as falling into universality
classes, each corresponding to a type of conformally invariant
behaviour.  Rather remarkably, the above statements hold true
whether we are dealing with a 2-dimensional classical statistical
system with some boundary condition imposed, or dealing with a
(1+1)-dimensional quantum system with some dynamical degrees of
freedom living on the boundary.  In fact, we already saw an example
of this in the single-channel Kondo problem.  The dynamical
impurity drops out of the description of the low-energy physics and
is replaced by a simple, scale-invariant boundary condition,
$\psi_L=-\psi_R$.

Precisely what is meant by a conformally invariant boundary
condition?  Without boundaries, conformal transformations are
analytic mappings of the complex plane: \begin{equation} z\equiv
\tau+ix,\end{equation} into itself:  \begin{equation} z\to
w(z).\end{equation}
 (Henceforth, we set the Fermi velocity, $v_F=1$.)  We may Taylor
expand an arbitrary conformal transformation around the origin:
\begin{equation} w(z) = \sum_0^\infty
a_nz^n,\label{CTexp}\end{equation} where the $a_n$'s are arbitrary
complex coefficients. They label the various generators of the
conformal group.  It is the fact that there is an infinite number
of generators (i.e. coefficients) which makes conformal invariance
so powerful in (1+1) dimensions.
 Now suppose that we have a boundary at $x=0$, the real axis.  At
best, we might hope to have invariance under all transformations
which leave the boundary fixed.  This implies the condition:
\begin{equation} w(\tau )^* = w(\tau ).\end{equation}  We see that
there is still an infinite number of generators, corresponding to
the $a_n$'s of Eq. (\ref{CTexp}) except that now we must impose the
conditions: \begin{equation} a_n^* = a_n.\end{equation}  We have
reduced the (still $\infty$) number of generators by a factor of
1/2.  The fact that there is still an $\infty$ number of
generators, even in the presence of a boundary, means that this
boundary conformal symmetry remains extremely powerful.

\begin{figure}
\epsfxsize=10 cm
\centerline{\epsffile{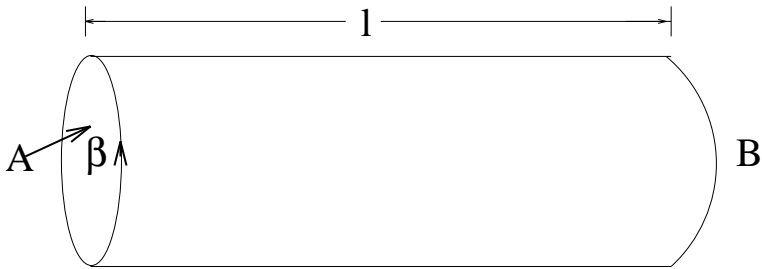}}
\caption{Cylinder of length $l$, circumference $\beta$ with
boundary conditions $A$ and $B$ at the two ends.}
\label{fig:cyl}
\end{figure}

To exploit this symmetry, following Cardy, it is very convenient to
consider a conformally invariant system defined on a cylinder of
circumference $\beta$ in the $\tau$-direction and length $l$ in the
$x$ direction, with conformally invariant boundary conditions $A$
and $B$ at the two ends.  [See Figure (\ref{fig:cyl}).] From the
quantum mechanical point of view, this corresponds to a finite
temperature, $T=1/\beta$.  The partition function for this system
is: \begin{equation} Z_{AB} = \hbox{tr}e^{-\beta
H^l_{AB}},\label{ZAB1}\end{equation} where we are careful to label
the Hamiltonian by the boundary conditions as well as the length of
the spatial interval, both of which help to determine the
spectrum.  Alternatively, we may make a modular transformation,
$\tau \leftrightarrow x$.  Now the spatial interval, of length,
$\beta$, is periodic.  We write the corresponding Hamiltonian as
$H^\beta_P$.  The system propagates for a time interval $l$ between
initial and final states $A$ and $B$.  Thus we may equally well
write: \begin{equation} Z_{AB} =
<A|e^{-lH^\beta_P}|B>.\label{ZAB2}\end{equation} Equating these two
expressions, Eq. (\ref{ZAB1}) and (\ref{ZAB2}) gives powerful
constraints which allow us to determine the conformally invariant
boundary conditions.

To proceed, we make a further weak assumption about the boundary
conditions of interest.  We assume that the momentum density
operator, $T-\bar T$ vanishes at the boundary.  This amounts to a
type of unitarity condition.  In the free fermion theory this
becomes: \begin{equation} \psi^{\dagger \alpha i}_L\psi_{L \alpha i}
(t,0)-\psi^{\dagger \alpha i}_R\psi_{R \alpha
i}(t,0)=0.\end{equation} Note that this is consistent with both
boundary conditions that occured in the one-channel Kondo problem:
$\psi_L=\pm \psi_R$.

Since $T(t,x)=T(t+x)$ and $\bar T(t,x)=\bar T(t-x)$, it follows that
\begin{equation} \bar T(t,x)=T(t,-x).\end{equation} i.e. we may
regard $\bar T$ as the analytic continuation of $T$ to the negative
axis.  Thus, as in our previous discussion, instead of working with
left and right movers on the half-line we may work with left-movers
only on the entire line.  Basically, the energy momentum density,
$T$ is unaware of the boundary condition.  Hence, in calculating
the spectrum of the system with boundary conditions $A$ and $B$
introduced above, we may regard the system as being defined
periodically on a torus of length $2l$ with left-movers only.  The
conformal towers of $T$ are unaffected by the boundary conditions,
$A$, $B$.  However, which conformal towers occur {\it does} depend
on these boundary conditions.  We introduce the characters of the
Virasoro algebra, for the various conformal towers:
\begin{equation} \chi_a (e^{-\pi \beta /l})\equiv \sum_ie^{-\beta
E^a_i(2l)},\end{equation} where $E^a_i(2l)$ are the energies in the
$a^{\hbox{th}}$ conformal tower for length $2l$.  i.e.:
\begin{equation} E^a_i(2l) = {\pi \over l}x_i^a-{\pi c\over 24
l},\end{equation} where the $x_i^a$'s correspond to the (left)
scaling dimensions of the operators in the theory and $c$ is the
conformal anomaly.  The spectrum of $H^l_{AB}$ can only consist of
some combination of these conformal towers.  i.e.: \begin{equation}
Z_{AB}=\sum_an^a_{AB}\chi_a(e^{-\pi \beta
/l}),\label{naAB}\end{equation} where the $n^a_{AB}$ are some
non-negative integers giving the multiplicity with which the
various conformal towers occur.  Importantly, only these
multiplicities depend on the boundary conditions, not the
characters, which are a property of the bulk left-moving system.
Thus, a specification of all possible multiplicities, $n^a_{AB}$
amounts to a specification of all possible boundary conditions
$A$.  The problem of specifying conformally invariant boundary
conditions has been reduced to determining sets of integers,
$n^a_{AB}$. For rational conformal field theories, where the number
of conformal towers is finite, only a finite number of integers
needs to be specified.

Now let us focus on the boundary states, $|A>$.  These must obey
the operator condition: \begin{equation} [T(x)-\bar T(x)]|A>=0 \ \
(\forall x).\end{equation}  Fourier transforming with respect to
$x$, this becomes: \begin{equation} [L_n-\bar
L_n]|A>=0.\end{equation}  This implies that all boundary states,
$|A>$ must be linear combinations of the ``Ishibashi
states'':\cite{Ishibashi} \begin{equation} |a> \equiv
\sum_m|a;m>\otimes \overline{|a;m>}.\label{Ishibashi}\end{equation}
Here $m$ labels all states in the $a^{th}$ conformal tower.  The
first and second factors in Eq. (\ref{Ishibashi}) refer to the left
and right-moving sectors of the Hilbert Space.  Thus we may write:
\begin{equation} |A> = \sum_a|a><a0|A>.\end{equation} Here,
\begin{equation} |a0>\equiv |a;0>\otimes
\overline{|a;0>}.\end{equation} (Note that while the states,
$|a;m>\otimes \overline{|b;n>}$ form a complete orthonormal set,
the Ishibashi states, $|a>$ do not have finite norm.)  Thus,
specification of boundary states is reduced to determining the
matrix elements, $<a0|A>$.  (For rational conformal field theories,
there is a finite number of such matrix elements.)  Thus the
partition function becomes: \begin{equation} Z_{AB} =
\sum_a<A|a0><a0|B><a|e^{-lH^\beta_P}|a>.\end{equation} From the
definition of the Ishibashi state, $|a>$ we see that:
\begin{equation}  <a|e^{-lH^\beta_P}|a>=\sum_me^{-2lE_m^a(\beta )
},\end{equation} the factor of 2 in the exponent arising from the
equal contribution to the energy from $T$ and $\bar T$.  This can be
written in terms of the characters: \begin{equation}
<a|e^{-lH^\beta_P}|a>=\chi_a (e^{-4\pi l/\beta}).\end{equation}

 We are now in a position to equate these two expressions for
$Z_{AB}$: \begin{equation} Z_{AB} =
\sum_a<A|a0><a0|B>\chi_a(e^{-4\pi
l/\beta})=\sum_an^a_{AB}\chi_a(e^{-\pi \beta
/l}).\label{ZAB12}\end{equation} This equation must be true for all
values of $l/\beta$.  It is very convenient to use the modular
transformation of the characters:\cite{Kac,Cardy2} \begin{equation}
\chi_a(e^{-\pi \beta /l})=\sum_bS^b_a\chi_b(e^{-4\pi
l/\beta}).\end{equation}  Here $S^a_b$ is known as the ``modular
S-matrix''.  (This name is rather unfortunate since this matrix has
no connection with the scattering-matrix.)  We thus obtain a set of
equations relating the multiplicities, $n^a_{AB}$ which determine
the spectrum for a pair of boundary conditions and the matrix
elements $<a0|A>$ determining the boundary states: \begin{equation}
\sum_bS^a_bn^b_{AB} = <A|a0><a0|B>.\label{Cardy}\end{equation} We
refer to these as Cardy's equations.  They basically allow a
determination of the boundary states and spectrum.

How do we go about constructing boundary states and multiplicities
which satisfy these equations?  Generally, boundary states
corresponding to trivial boundary conditions can be found by
inspection.  i.e., given $n^b_{AA}$ we can find $<a|A>$.  We can
then generate new (sometimes non-trivial) boundary states by {\it
fusion}.  i.e. given any conformal tower, $c$, we can obtain a new
boundary state $|B>$ and new spectrum $n^a_{AB}$ from the ``fusion
rule coefficients'', $N^c_{ab}$.  These non-negative integers are
defined by the operator product expansion (OPE) for (chiral)
primary operators, $\phi_a$.  In general the (OPE) of $\phi_a$ with
$\phi_b$ contains the operator $\phi_c$ $N^c_{ab}$ times.  In
simple cases, such as occur in the Kondo problem, the $N^c_{ab}$'s
are all 0 or 1.  In the case of $SU(2)_k$, which will be relevant
for the Kondo problem, the OPE
 is:\cite{Zamolodchikov1,Gepner} \begin{equation} j \otimes j' =
|j-j'|, |j-j'|+1, |j-j'|+2, \ldots ,\hbox{min} \{ j+j', k-j-j'\}
.\end{equation}   Note that this generalizes the ordinary angular
momentum addition rules
 in a way which is consistent with the conformal tower structure of
the theories (i.e. the fact that primaries only exist with $j\leq
k/2$).
 Thus,  \begin{eqnarray} N^{j''}_{jj'}&=&1\ \  (|j-j'|\leq j'' \leq
\hbox{min} \{ j+j', k-j-j'\} \nonumber \\ &=&0\ \
\hbox{otherwise}.\label{KMfusion}\end{eqnarray}

The new boundary state, $|B>$, and multiplicities obtained by
fusion with the conformal tower $c$ are given by: \begin{eqnarray}
<a0|B> &=& <a0|A>{S^a_c\over S^a_0}\nonumber \\ n^a_{AB} &=&
\sum_bN^a_{bc}n^b_{AA}.\label{fusion}\end{eqnarray} Here $0$ labels
the conformal tower of the identity operator.  Importantly, the new
boundary state and multiplicities so obtained, obey Cardy's
equation.  The right-hand side of Eq.  (\ref{Cardy}) becomes:
\begin{equation} <A|a0><a0|B> = <A|a0><a0|A>{S^a_c\over
S^a_0}.\end{equation}  The left-hand side becomes: \begin{equation}
\sum_bS^a_bn^b_{AB} = \sum_{b,d}S^a_bN^b_{dc}n^d_{AA}.\end{equation}
We now use a remarkable identity relating the modular S-matrix to
the fusion rule coefficients, known as the Verlinde
formula:\cite{Verlinde} \begin{equation}
\sum_bS^a_bN^b_{dc}={S^a_dS^a_c\over S^a_0}.\end{equation}  This
gives: \begin{equation} \sum_bS^a_bn^b_{AB} = {S^a_c\over
S^a_0}\sum_dS^a_dn^d_{AA}={S^a_c\over
S^a_0}<A|a0><a0|A>=<A|a0><a0|B>,\end{equation} proving that fusion
does indeed give a new solution of Cardy's equations. The
multiplicities, $n^a_{BB}$ are given by double fusion:
\begin{equation} n^a_{BB} =
\sum_{b,d}N^a_{bc}N^b_{dc}n^d_{AA}.\end{equation}   [Recall that
$|B>$ is obtained from $|A>$ by fusion with the primary operator
$c$.]  It can be checked that the Cardy equation with $A=B$ is then
obeyed.  It is expected that, in general, we can generate a
complete set of boundary states from an appropriate reference state
by fusion with all possible conformal towers.

\section{Boundary Conformal Field Theory Results on the
Multi-Channel Kondo Effect} \subsection{Fusion and the Finite-Size
Spectrum} We are now in a position to bring to bear the full power
of boundary conformal field theory on the Kondo problem.  By the
arguments at the beginning of Sec. V, we expect that the infrared
fixed points describing the low-T properties of the Kondo
Hamiltonian correspond to conformally invariant boundary conditions
on free fermions.  We might also expect that we could determine
these boundary conditions and corresponding boundary states by
fusion with appropriate operators beginning from some convenient,
trivial, reference state.

We actually already saw a simple example of this in Sec. III in the
single channel, $s=1/2$, Kondo problem.  There we observed that the
free fermion spectrum, with convenient boundary conditions could be
written: \begin{equation} (0,\hbox{even})\oplus
(1/2,\hbox{odd}).\end{equation} Here $0$ and $1/2$ label the
$SU(2)_1$ KM conformal towers in the spin sector, while ``even''
and ``odd'' label the conformal towers in the charge sector.  We
argued that, after screening of the impurity spin, the infrared
fixed point was described by free fermions with a $\pi /2$ phase
shift, corresponding to a spectrum: \begin{equation}
(1/2,\hbox{even})\oplus (0,\hbox{odd}).\end{equation} The change in
the spectrum corresponds to the interchange of $SU(2)_1$ conformal
towers: \begin{equation} 0 \leftrightarrow 1/2.\end{equation} This
indeed corresponds to fusion, with the spin-1/2 primary field of
the WZW model.  To see this note that the fusion rules for $SU(2)_1$
are simply [from  Eq. (\ref{KMfusion})]: \begin{eqnarray} 0\otimes
{1\over 2} &=& {1\over 2}\nonumber \\ {1\over 2}\otimes {1\over 2}
&=& 0.\end{eqnarray}  Thus for an $s=1/2$ impurity, the infrared
fixed point is given by fusion with the $j=1/2$ primary.  This is
related to our completing the square argument.  The new currents at
the infrared fixed point, $\vec {\cal J}$, are related to the old
ones, $\vec J$, by: \begin{equation} \vec {\cal J}_n = \vec J_n +
\vec S.\label{square}\end{equation} If $\vec J$ and $\vec {\cal J}$
were ordinary spin operators, then the new spectrum would be given
by the ordinary angular momentum addition rules.  In the case at
hand, where $\vec J$ and $\vec {\cal J}$ are KM current operators,
it is plausible that the spectrum is given by fusion with the
spin-s representation, generalizing the ordinary angular momentum
addition rules in  a way which is consistent with the structure of
the KM CFT.  In particular, Eq. (\ref{square}) implies, for
half-integer $s$, that states of integer
 total spin are mapped into states of half-integer total spin, and
vice versa, a property which follows from fusion with $j=1/2$.

This immediately suggests a way of determining the boundary
condition for arbitrary number of channels, k and impurity spin
magnitude, s: fusion with spin-s.  Actually, while this is possible
for $s\leq k/2$, corresponding to exact or overscreening, it is not
possible in the underscreened case since there is no spin-s primary
with which to fuse for $s>k/2$.  Instead, in the underscreened
case, we assume fusion with the maximal possible spin, namely
$k/2$.  This seems to correspond to the (in this case stable)
strong coupling fixed point described in Sec. III.  $k/2$ electrons
partially screen the impurity.  The fact that further screening is
not possible is related to Fermi statistics.  The maximal possible
conduction electron spin state at the origin, for k channels is
k/2.  This is also essentially the reason why there are no
primaries with larger spin, as can be seen from the corresponding
bosonization of free fermions.  We reiterate this essential point:
{\it The infrared fixed point in the k-channel spin-s Kondo problem
is given by fusion with the spin-s primary for $s\leq k/2$ or with
the spin $k/2$ primary for $s>k/2$.}  We have referred to this as
the ``fusion rules hypothesis''. If the general assumption that the
infrared fixed point should be described by a conformally invariant
boundary condition is accepted, then this hypothesis starts to seem
very plausible.  The general method for generating new boundary
conditions is by fusion.  Since the Kondo interaction appears
entirely in the spin sector of the theory we should expect that the
fusion occurs in that sector.  The current redefinition $\vec J \to
\vec {\cal J}$ and various self-consistency checks all point
towards this particular set of fusions.

An immediate way of checking the fusion rule hypothesis, and more
generally the applicability of the boundary CFT framework to this
problem, is to work out in detail the finite size spectrum for a few
values of k and s and compare with spectra obtained by numerical
methods.

Let us first consider the exactly screened and underscreened cases,
$s\geq k/2$, where fusion occurs with the spin k/2 primary.  In this
case the fusion rules are simply: \begin{equation} j\otimes {k\over
2} = {k\over 2}-j.\end{equation}  Each conformal tower is mapped
into a unique conformal tower.  It can be shown that this gives the
free fermion spectrum with a $\pi /2$ phase shift.\cite{Affleck2}

We demonstrate the case k=2, s=1 in Tables (\ref{tab:antiper}) and
(\ref{tab:per}).  Let us start with antiperiodic boundary
conditions in the left-moving formalism: \begin{equation} \psi_L(l)
= -\psi_L(-l).\end{equation}  Let us express this free fermion
spectrum, for 2 spin components and 2 channels, in terms of
products of conformal towers in the charge, spin and flavour
sectors.  In this case, the flavour sector corresponds to $SU(2)_2$
as does the flavour sector.  We will refer to the corresponding
quantum numbers as j for ordinary spin and $j_f$ for flavour (or
``pseudo-spin'').  We need the energy of the ``highest weight
state'' (i.e. groundstate) of each conformal tower.  For the
Kac-Moody conformal towers, the highest weight state transforming
under the representation R of the group G at level k has
energy:\cite{Knizhnik} \begin{equation} E_R={\pi \over l}{C_R\over
k+C_A},\end{equation} where $C_R$ is the quadratic Casimir in the R
representation and $A$ refers to the fundamental representation.
For the case of SU(2) the representations are labelled by their
spin, j and the Casimirs are: \begin{equation} C_j =
j(j+1).\end{equation} We also need the energy for the charge
sector.  These can be worked out by generalizing the method used in
the k=1 case in Sec. II.  The energy for the lowest charge $Q$
excitation, for each species of fermion is: \begin{equation} E={\pi
\over l}{Q^2\over 2},\end{equation} as shown in Eq.
(\ref{1spectrum}).  Altogether we obtain 4 terms like this for the
4 species of fermions (2 spin $\times$ 2 flavours).  We can express
the total energy in terms of the total charge: \begin{equation}
Q\equiv Q_{11}+Q_{12}+Q_{21}+Q_{22},\end{equation} and various
difference variables.  This gives: \begin{equation} E={\pi \over
l}{Q^2\over 8} + ...\end{equation} This gives the energy of the
charge $Q$ primary.  In addition to the energy of the primary state
we obtain additional terms in the energy corresponding to the
excitation level in the charge, spin and flavour conformal towers:
$n_Q$, $n_s$ and $n_f$. These are non-negative integers.
Altogether, we may write the energy of any state as:
\begin{equation} E={\pi \over l}\left[ {Q^2\over 8}+{j(j+1)\over 4}
+ {j_f(j_f+1)\over 4}+ n_Q+n_s+n_f\right]
.\label{Ek=2}\end{equation}  For primary states, $n_Q=n_s=n_f=0$.
$Q$ must be integer and $j$ and $j_f$ must be integer or
half-integer.  The allowed combinations of $Q$, $j$ and $j_f$ are
what we refer to as ``gluing conditions''.  They depend on the
boundary conditions.  For antiperiodic boundary conditions, the
allowed fermion momenta are  \begin{equation} k = \pi
(n+1/2)/l.\end{equation} The corresponding energy levels are drawn
in Figure (\ref{fig:anti}).  Note that the groundstate is unique.
It has $j=j_f=Q=0$.  Thus we must include the corresponding product
of conformal towers in the spectrum.  The single particle or single
hole excitation has $j=j_f=1/2$ and $Q=\pm 1$.  The energy is:
\begin{equation} E={\pi \over l}\left[{1\over 8} + {3/4\over
4}+{3/4\over 4}\right]={\pi \over l}{1\over 2},\end{equation} the
right value.   2-particle excitations have $lE/\pi = 1$, $Q=2$ and
either have $j=1$, $j_f=0$ or $j=0$, $j_f=1$, due to Fermi
statistics.  Again the energy is given correctly by Eq.
(\ref{Ek=2}).  The lowest energy particle-hole excitations, with
 $lE/\pi = 1$, have $Q=0$ and various values of $j$ and $j_f$.  In
particular, they can have $j=j_f=1$.   It turns out that these
excitations are Kac-Moody primaries.  Again the energy is given
correctly by Eq. (\ref{Ek=2}).  It can be shown\cite{Altshuler} that
these are all conformal towers that occur, except that we must allow
arbitrary values of $Q$, mod 4.  These conformal towers are
summarized in Table (\ref{tab:antiper}).
\begin{table}\caption{Conformal towers appearing in the $k=2$ free
fermion spectrum with anti-periodic boundary conditions.}
\label{tab:antiper}\begin{tabular}{llll} $Q$ (mod 4)&
$j$&$j_f$&$(El/\pi )_{\hbox{min}}$\\ \tableline 0&0&0&0\\ 0&1&1&1\\
$\pm$1&1/2&1/2&1/2\\ 2&0&1&1\\ 2&1&0&1\\ \end{tabular}\end{table}
Now consider fusion with the $j=1$ primary.  This has the effect of
shuffling the spin conformal towers in the following way:
\begin{eqnarray} 0 &\rightarrow & 1\nonumber \\ 1/2 & \rightarrow
&1/2 \nonumber \\ 1 & \rightarrow 0.\end{eqnarray}  The spectrum of
Table (\ref{tab:antiper}) goes into that of Table (\ref{tab:per})
under this shuffling.  It can be checked that this corresponds to
free fermions with a $\pi /2$ phase shift, i.e. periodic boundary
conditions, or a shift of the Fermi energy by 1/2 a level spacing,
drawn in Figure (\ref{fig:per}).  Now note that the groundstate is
$2^4=16$-fold degenerate, since the zero-energy level may be filled
or empty for each species of fermion.  The charge $,Q$, in Table
(\ref{tab:per}) is now measured relative to the symmetric case
where 2 of these levels are filled and 2 are empty.  Also note that
if make the replacement: \begin{equation} Q \to Q-2,\end{equation}
we get back the previous spectrum of Table (\ref{tab:antiper}).
Making this replacement in Eq. (\ref{Ek=2}), we obtain:
\begin{equation}{ El\over \pi} \to {El\over \pi}+{Q\over
2}\end{equation} (ignoring a constant).  This corresponds to
shifting the Fermi energy by 1/2-spacing; i.e. a $\pi /2$ phase
shift. \begin{table}\caption{Conformal towers appearing in the
$k=2$ free fermion spectrum with periodic boundary conditions,
obtained from the  Table I by fusion with
$j=1$.}\label{tab:per}\begin{tabular}{llll} $Q$ (mod 4)&
$j$&$j_f$&$(El/\pi )_{\hbox{min}}$\\ \tableline 0&1&0&1/2\\
0&0&1&1/2\\ $\pm$1&1/2&1/2&1/2\\ 2&1&1&3/2\\ 2&0&0&1/2\\
\end{tabular}\end{table}

In the overscreened case the fusion rules are more interesting.
They lead to spectra which cannot be obtained by applying any simple
linear boundary conditions to the free fermions.  Thus we may refer
to these as non-Fermi liquid fixed points.  It might be possible to
find some kind of non-linear description of the boundary conditions
in this case.  But note that a non-linear boundary condition
effectively introduces an interaction into the theory at the
boundary.  A boundary condition quadratic in the fermion fields
might induce an additional condition quartic in fields, etc.  Thus
specification of non-linear boundary conditions could be very
difficult.  Cardy's formalism cleverly sidesteps this problem by the
device of focussing on the boundary {\it states} instead of boundary
conditions, and providing a method (fusion) for producing these
boundary states.  As was stated above, and we will continue to see
in what follows, knowledge of the boundary states will determine all
physical properties of the theory so nothing is lost by using this
abstract description of the boundary condition.

For the $k=2$, $s=1/2$ example, the fusion rules give:
\begin{eqnarray} 0 &\to & 1/2 \nonumber \\ 1/2 &\to & 0 \oplus 1
\nonumber \\ 1 &\to & 1/2.\end{eqnarray} Now we get a larger number
of conformal towers in the spectrum with this boundary condition,
shown in Table (\ref{tab:1/2fus}).  We have shifted the groundstate
energy to 0, in this Table.  Note that energies $El/\pi = 1/8$ and
$5/8$ now occur.  These do not correspond to any possible linear
boundary conditions on free fermions.  Note in particular that, due
to particle-hole symmetry, only phase shifts of 0 or $\pi /2$ are
allowed.  These give half-integer energies, as we saw above and in
Tables (\ref{tab:antiper}) and (\ref{tab:per}).

\begin{table}\caption{Conformal towers appearing in the $k=2$ free
fermion spectrum after fusion with
$j=1/2$.}\label{tab:1/2fus}\begin{tabular}{llll}  $Q$ (mod
4)&$j$&$j_f$&$(El/\pi )_{\hbox{min}}$\\ \tableline  0&1/2&0&0\\
0&1/2&1&1/2\\ $\pm$ 1&0&1/2&1/8 \\ $\pm$ 1&1&1/2&5/8\\
 2&1/2&0&1/2\\ 2&1/2&1&1 \end{tabular}\end{table} This spectrum was
compared with numerical work on the $k=2$, $s=1/2$ Kondo effect and
the agreement was excellent (to within 5\% for several of the
lowest energy states).\cite{Affleck5}  This provides  evidence that
the fusion rule hypothesis is correct in the overscreened case.
\subsection{Impurity Entropy} We define the impurity entropy as:
\begin{equation} S_{\hbox{imp}}(T) \equiv lim_{l\to
\infty}[S(l,T)-S_0(l,T)],\label{Simpdef}\end{equation} where
$S_0(l,T)$ is the free fermion entropy, proportional to $l$, in the
absence of the impurity.  We will find an interesting, non-zero
value for $S_{\hbox{imp}}(0)$.  Note that, for zero Kondo coupling,
$S_{\hbox{imp}}=\ln [s(s+1)],$ simply reflecting the groundstate
degeneracy of the free spin.  In the case of exact screening,
($k=2s$), $S_{\hbox{imp}}(0)=0$.  For underscreening,
\begin{equation} S_{\hbox{imp}}(0) = \ln [s'(s'+1)],\end{equation}
where $s'\equiv s-k/2$.  What happens for overscreening?
Surprisingly, we will obtain, in general, the log of a non-integer,
implying a sort of ``non-integer groundstate degeneracy''.

To proceed, we show how to calculate $S_{\hbox{imp}}(0)$ from the
boundary state.  All calculations are done in the scaling limit,
ignoring irrelevant operators, so that  $S_{\hbox{imp}}(T)$ is a
constant, independent of $T$, and characterizing the particular
boundary condition.  It is important, however, that we take the
limit $l\to \infty$ first, as specified in Eq. (\ref{Simpdef}), at
fixed, non-zero $T$.  i.e. we are interested in the limit, $l/\beta
\to \infty$.  Thus it is convenient to use the first expression for
the partition function, $Z_{AB}$ in Eq. (\ref{ZAB2}):
\begin{equation} Z_{AB} = \sum_a<A|a0><a0|B>\chi_a(e^{-4\pi
l/\beta})\to e^{\pi lc/6\beta}<A|00><00|B>.\end{equation}  Here
$|00>$ labels the groundstate in the conformal tower of the
identity operator.  $c$ is the conformal anomaly.   Thus the free
energy is: \begin{equation} F_{AB} = -\pi cT^2l/6-T\ln
<A|00><00|B>.\end{equation} The first term gives the specific heat:
\begin{equation} C=\pi cTl/3\end{equation} and the second gives the
impurity entropy: \begin{equation} S_{\hbox{imp}} = \ln
<A|00><00|B>.\end{equation}  This is a sum of contributions from
the two boundaries,  \begin{equation}
S_{\hbox{imp}}=S_A+S_B.\end{equation}  Thus we see that the
``groundstate degeneracy'' $g_A$, associated with boundary
condition A is: \begin{equation} \exp [S_{\hbox{imp}A}] =
<A|00>\equiv g_A.\end{equation}  Here we have used our freedom to
choose the phase of the boundary state so that $g_A>0$.  For our
original, anti-periodic, boundary condition, $g=0$. For the Kondo
problem we expect the low T impurity entropy to be given by the
value at the infrared fixed point.  Since this is obtained by
fusion with the spin-s (or k/2) operator, we obtain from Eq.
(\ref{fusion}), \begin{equation} g = {S^0_s\over
S^0_0}.\end{equation}  The modular S-matrix for $SU(2)_k$
is:\cite{Altshuler,Kac} \begin{equation} S^j_{j'} (k) =
\sqrt{2\over 2+k}\sin \left[{\pi (2j+1)(2j'+1)\over
2+k}\right],\end{equation} so \begin{equation} g(s,k) = {\sin [\pi
(2s+1)/(2+k)]\over \sin [\pi /(2+k)]}.
\label{Smatrix}\end{equation}  This formula agrees exactly with the
Bethe ansatz  result.\cite{Tsvelik} This formula has various
interesting properties.  Recall that in the case of exact or
underscreening ($s\geq k/2$) we must replace $s$ by $k/2$ in this
formula, in which case it reduces to 1.  Thus the groundstate
degeneracy is 1 for exact screening.  For underscreening we must
multiply $g$ by $(2s'+1)$ to account for the decoupled, partially
screened impurity.  Note that, in the overscreened case, where $s<
k/2$, we have: \begin{equation} {1\over 2+k}<{2s+1\over
2+k}<1-{1\over 2+k},\end{equation} so $g>1$. In the case $k\to
\infty$ with s held fixed, $g\to 2s+1$, i.e. the entropy of the
impurity spin is hardly reduced at all by the Kondo interaction,
corresponding to the fact that the critical point occurs at weak
coupling.  In general, for underscreening: \begin{equation}
1<g<2s+1.\end{equation} i.e. the free spin entropy is somewhat
reduced, but not completely eliminated.  Furthermore, $g$ is not,
in general, an integer. For instance, for $k=2$ and $s=1/2$,
$g=\sqrt{2}$. Thus we may say that there is a non-integer
``groundstate degeneracy''.  Note that in all cases the groundstate
degeneracy is reduced under renormalization from the zero Kondo
coupling fixed point to the infrared stable fixed point.  This is a
special case of what we believe to be a general result: {\it the
groundstate degeneracy always decreases under renormalization.}
This appears to be related to Zamolodchikov's
c-theorem\cite{Zamolodchikov2} which states that the conformal
anomaly parameter, c, always decreases under renormalization.  The
intuitive explanation of the c-theorem is that, as we probe lower
energy scales, degrees of freedom which appeared approximately
massless start to exhibit a mass.  This freezes out their
contribution to the specific heat, the slope of which can be taken
as the definition of c.  In the case of the ``g-theorem'' the
intuitive explanation is that, as we probe lower energy scales,
approximately degenerate levels of impurities exhibit small
splittings, reducing the degeneracy.

So far, only a perturbative proof of the g-theorem has been
given.\cite{Affleck6} It is completely analogous to a perturbative
proof of the c-theorem given by Cardy and Ludwig,\cite{Ludwig3}
independently of Zamolodchikov's more general proof.  For the
g-theorem proof, we consider perturbing around a boundary CFT fixed
point with a barely relevant boundary operator.  i.e. the action is:
\begin{equation} S = S_0-\lambda \int_0^\beta d\tau \phi (0,\tau
),\end{equation} where $\phi$ has dimension $1-y$ with $0<y<<1$.
The $\beta$-function has the form: \begin{equation} \beta =
y\lambda -b\lambda^2,\end{equation} for some constant, b.  There is
a nearby fixed point at: \begin{equation}
\lambda_c=y/b.\end{equation}  It is possible to calculate the small
change in $g$ using renormalization group improved perturbation
theory.  This gives: \begin{equation} \delta g/g = -\pi^2y^3/3b^2
<0.\end{equation} \subsection{Boundary Green's Functions: Two-Point
Functions, T=0 Resistivity} In this sub-section we explain the
basic concepts for calculation of
 Green's functions in the presence of a conformally invariant
boundary condition.\cite{Cardy3}
  We then work out the case of two-point funtions in detail.
Finally we show how this gives information about the Kondo
problem.

\begin{figure}
\epsfxsize=10 cm
\centerline{\epsffile{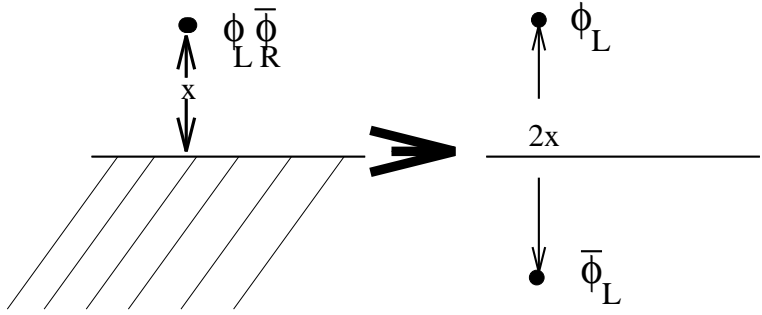}}
\caption{A local operator becomes effectively bilocal
in the presence of a boundary.}
\label{fig:1to2}
\end{figure}

The most important point is the consequence of the identification of
left and right-moving sectors, discussed in Sec. V.  In general, in
the bulk theory, a typical local operator is a product of left and
right-moving factors: \begin{equation} \phi (x) = \phi_L(x)\bar
\phi_R(x).\end{equation} Here $x$ is the spatial co-ordinate; we
suppress the time-dependence.  However, in the presence of a
boundary, we use: \begin{equation} \bar \phi_R(x) = \bar
\phi_L(-x).\end{equation}  Thus a local operator with left and
right-moving factors becomes a bilocal operator with only
left-moving factors: \begin{equation} \phi (x) \to \phi_L(x)\bar
\phi_L(-x).\end{equation}  [See Figure (\ref{fig:1to2}).] Thus a
one-point function becomes a two-point function, two-point becomes
four-point etc.

Henceforth, the number of points in the Green's function will refer
to the larger number after this doubling due to the identification
of left with right.  In the remainder of this sub-section we show
how to calculate boundary two-point functions.  Four point
functions are discussed in the next sub-section.

Our bulk operators are normally defined so that $<\phi
(x)>_{\hbox{bulk}}=0$.  For a semi-infinite plane with a boundary,
this one-point function essentially becomes a 2-point function
which may have a non-zero value: \begin{equation} <\phi
(x)>_A=<\phi_L(x)\bar \phi_L(-x)>_A = {C_A\over (2x)^{2d}}.
\label{2pt}\end{equation} Here $d$ is the scaling dimension of
$\phi_L$, which does not depend on the boundary condition, A.  On
the other hand, the coefficient, $C_A$ does depend on the boundary
condition.  Following Cardy and Lewellen,\cite{Cardy2} we may
calculate $C_A$ in terms of the boundary state $|A>$.  We assume
that $\phi$ is a primary field.

\begin{figure}
\epsfxsize=10 cm
\centerline{\epsffile{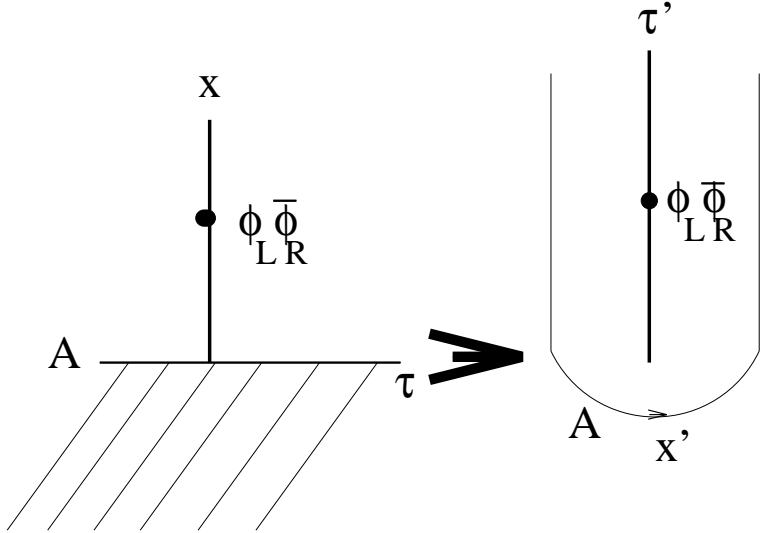}}
\caption{Conformal mapping of a two-point function
{}from the semi-infinite plane to the semi-infinite
cylinder.}
\label{fig:pltocyl}
\end{figure}

This is done by making a conformal mapping from the semi-infinite
cylinder  to the semi-infinite plane: \begin{equation} z=i\tanh
{\pi w\over \beta}.\label{cylinder}\end{equation}  Writing:
\begin{eqnarray} z&=&\tau + ix \nonumber \\ w&=& \tau
'+ix',\end{eqnarray} we see that, as $x'$ goes from $-\beta /2$ to
$\beta /2$, $\tau $ goes from $-\infty$ to $\infty$. [See Figure
(\ref{fig:pltocyl}).] The semi-infinite plane is $x>0$ and the
semi-infinite cylinder is $\tau '>0$.  Note that we are regarding
$x'$ (space) as the periodic variable on the cylinder.  i.e. we are
calculating imaginary-time propagation from the boundary state
$|A>$.   We will  calculate the one-point function on the cylinder,
first directly, then by obtaining it from  the half-plane by
conformal mapping.  To obtain the correlation function on the {\it
infinite} half-cylinder, it is convenient to take the limit of a
finite cylinder, with boundary state $|00>$ (the highest weight
state of the identity conformal tower) at the other end.  Thus, on
the infinite cylinder, \begin{equation} <\phi (\tau ',0)>_A =
lim_{T\to \infty}{<00|e^{-(T-\tau ' )H^\beta_P}\phi (0,0)e^{-\tau '
H_P^\beta}|A>\over <00|e^{-TH_P^\beta}|A>}.\end{equation}
 Now we insert a complete set of states between $\phi$ and $|A>$.
Since $\phi$ is a primary field, $\phi |00>$ gives only a sum of
states in the conformal tower of $\phi$.  For convenience, we also
consider  the limit, $\tau '\to \infty$, so that only the highest
weight state, $|\phi 0>$ survives. Thus, \begin{equation} <\phi
(\tau ')>_A\to_{\tau '\to \infty} {<00|\phi |\phi 0><\phi 0|A>\over
<00|A>}e^{-(2\pi /\beta )2d\tau '}.\end{equation} Note that
\begin{equation} E_\phi = (2\pi /\beta )2d.\end{equation} We need
the matrix element $<00|\phi |\phi 0>$, for the periodic
Hamiltonian $H^\beta_P$.  This can be obtained from the Green's
function $<00|\phi (\tau_1')\phi (\tau_2')|00>$, arising from a
calculation on the infinite cylinder of radius
$\beta$.\cite{Cardy2}  This can be obtained by a conformal mapping
{}from the infinite plane, giving: \begin{eqnarray} <00|\phi
(\tau_1')\phi (\tau_2')|00>&=&\left[ {\beta \over \pi}\sinh {\pi
\over \beta}(\tau_1'- \tau_2' )\right]^{-4d}\nonumber \\
&\to&_{\tau_1' - \tau_2'\to \infty}\left({2\pi \over \beta
}\right)^{4d}e^{-4\pi d(\tau_1'-\tau_2')/\beta} \nonumber \\
&=&|<00|\phi |\phi 0>|^2e^{-E_\phi
(\tau_1'-\tau_2')}.\end{eqnarray}  Thus we obtain: \begin{equation}
<00|\phi |\phi >=\left({2\pi \over
\beta}\right)^{2d}.\end{equation}  Thus the desired one-point
function, on the half-cylinder is: \begin{equation} <\phi (\tau '
)>_A \to_{\tau '\to \infty} \left({2\pi \over
\beta}\right)^{2d}{<\phi 0|A>\over <00|A>}e^{-(2\pi /\beta )2d\tau
'}.\label{1pthalf1}\end{equation}

Now let us repeat this calculation, by conformal  transformation
{}from the semi-infinite plane.  As argued above, the result on the
half-plane takes the form: \begin{equation} <\phi (x)>_A={C_A\over
(2x)^{2d}}.\end{equation} We now obtain the result on the
half-cylinder by the conformal transformation of Eq.
(\ref{cylinder}).  This gives: \begin{eqnarray} <\phi (\tau
)>_A^{\hbox{1/2-cylinder}}&=&|{dz\over dw}|^{2d}<\phi
(x)>_{A}^{\hbox{plane}}\nonumber \\ &=&\left({\pi \over
\beta}\hbox{sech}^2{\pi \tau '\over \beta}\right)^{2d}{C_A\over
(2x)^{2d}}\nonumber \\ &=&C_A\left({\pi \over
\beta}{\hbox{sech}^2\pi \tau '/\beta \over 2\tanh \pi \tau
'/\beta}\right)^{2d}\nonumber \\ &\to & C_A \left({2\pi \over
\beta}e^{-2\pi \tau '
/\beta}\right)^{2d}.\label{1pthalf2}\end{eqnarray}  Thus, equating
Eq. (\ref{1pthalf1}) and (\ref{1pthalf2}), we finally obtain the
desired formula for the coefficient $C_A$, defined in Eq.
(\ref{2pt}): \begin{equation} C_A = {<\phi 0|A >\over
<00|A>}.\end{equation} If $|A>$ is obtained by fusion with primary
operator $c$ from some reference state $|F>$ (standing for free),
then: \begin{equation} C_A=C_A^{\hbox{free}}{S^\phi_c/S^\phi_0\over
S^0_c/S^0_0}.\label{C_Afusion}\end{equation} Eq. (\ref{2pt}) can be
immediately generalized to the case where the left and right
factors occur at different points in the upper half-plane:
\begin{equation} <\phi_L(z)\bar \phi_R(z')>_A={C_A\over [(\tau
-\tau ')+i(x+x')]^{2d}}.\end{equation}

An important application of this formula in the Kondo problem is to
the single fermion Green's function,
$<\psi_L(z)^\dagger\psi_R(z')>$.  In the case of  periodic
boundary conditions: \begin{eqnarray} <\psi^{\dagger
i\alpha}_L(x)\psi_{Rj\beta}(x)>_{\hbox{free}} &= &
 <\psi^{\dagger i\alpha}_L(x)\psi_{Lj\beta}(-x)>\nonumber \\
&=&{\delta^i_j\delta ^\alpha_\beta \over 2x}.\end{eqnarray}  We may
obtain the one-point function at the Kondo fixed point by fusion.
The fermion operator can be written as a product of spin flavour
and charge operators: \begin{equation} \psi_{Lj\alpha} \propto
g_\alpha h_je^{i\sqrt{2\pi /k}\phi}. \label{bosonizek}\end{equation}
Here $g$ and $h$ are the left moving factors of the  primary fields
of the WZW models, transforming under the fundamental
representation.  i.e. $g$ has s=1/2.  We may use Eq.
(\ref{C_Afusion}).  The Kondo boundary condition is obtained by
fusion with the spin-s conformal tower, where s is the spin of the
impurity.  Thus: \begin{eqnarray} <\psi^{\dagger
i\alpha}_L(x)\psi_{Rj\beta}(x)>^{\hbox{Kondo}}_{s}
&=&{\delta^i_j\delta ^\alpha_\beta \over
2x}{S^{1/2}_s/S^{1/2}_0\over S^0_s/S^0_0}\nonumber \\ &=&{\cos [\pi
(2s+1)/(2+k)]\over \cos [\pi /(2+k)]} {\delta^i_j\delta
^\alpha_\beta \over 2x}\label{1pGF}\end{eqnarray}  Here we have
used the SU(2) modular S-matrix given in Eq. (\ref{Smatrix}) .
There are several interesting points to notice about this formula.
First of all, we see that for the case of exact or underscreening,
where $s=k/2$, the cosine in the numerator in Eq. (\ref{1pGF})
becomes: \begin{equation} \cos \pi (k+1)/(k+2) = -\cos
\pi/(k+2).\end{equation}  Thus: \begin{equation}
<\psi^\dagger_L\psi_R>^{\hbox{Kondo}}_{k/2} = -
 <\psi^\dagger_L\psi_R>_{\hbox{free}}.\end{equation}  As expected,
this corresponds to a $\pi /2$ phase shift: \begin{equation}
\psi_R(0) = e^{2i\delta}\psi_L(0)=-\psi_L(0).\end{equation} In
general, we may define a one-particle into one-particle S-matrix
element , $S^{(1)}$ by: \begin{equation}  <\psi^\dagger_L\psi_R> =
S^{(1)} <\psi^\dagger_L\psi_R>_{\hbox{free}},\end{equation} with:
\begin{equation} S^{(1)} = {\cos [\pi (2s+1)/(2+k)]\over \cos [\pi
/(2+k)]}.\label{S(1)}\end{equation} We see that in the overscreened
case, $|S^{(1)}|<1$.  $S^{(1)}$ is the matrix element, at the Fermi
energy, for a single electron to scatter off the impurity into a
single electron.  In the case where $|S^{(1)}|=1$, we see, by
unitarity, that there is zero probability for a single electron to
scatter into anything but a single electron at the Fermi energy, at
zero temperature.  This is precisely the starting point for
Landau's Fermi liquid theory.  Thus we refer to such cases as Fermi
liquid boundary conditions.  In the overscreened case, where
$|S^{(1)}|<1$, this inelastic scattering probability is non-zero so
we have non-Fermi liquid boundary conditions.  It is interesting to
note that in the large k limit (with s held fixed),  $S^{(1)}\to
1$, corresponding to the fixed point occuring at weak coupling. We
also note that, for $k=2$, $s=1/2$, $S^{(1)}=0$.  This is related
to the symmetry between the zero coupling ($S^{(1)}=1$) and
infinite coupling ($S^{(1)}=-1$) fixed point, mentioned above.  In
this case the one particle to one particle scattering rate vanishes!

 {}From the electron self-energy we can obtain the lifetime and
hence the resistivity for a dilute array of impurities by the Kubo
formula\cite{Affleck6} giving a $T=0$ resistivity: \begin{equation}
\rho (0) = {3n_i\over k\pi (e\nu v_F)^2}\left[ {1-S^{(1)}\over
2}\right] .\end{equation}  Here $n_i$ is the impurity density.  The
first factor is the ``unitary limit''.  i.e. this is the largest
possible resistivity that can occur for a dilute array of
non-magnetic impurities.  It is only realised in the Kondo problem
in the Fermi liquid case (exact or overscreened).  Otherwise the
resistivity is reduced by the factor $\left[ {1-S^{(1)}\over
 2}\right]$, which goes to zero for large k. \subsection{Four Point
Boundary Green's Functions, Spin-Density Green's Function} In this
sub-section we sketch how four-point functions of chiral operators
(or two-point functions of non-chiral ones) are calculated in the
presence of a boundary, using the particular example of the fermion
four-point function with the Kondo boundary condition. For more
details, see Ref. (\onlinecite{Ludwig2}). Similarly to the case of
the two-point function, we will find that the four-point function is
determined by bulk properties up to a
 constant which can be expressed in terms of matrix elements
involving the boundary state.

We consider the Green's function: \begin{equation} G\equiv
<\psi_{L\alpha i}(z_1)\psi_L^{\dagger \bar \beta \bar
j}(z_2)\psi_{R\beta j}(z_3)\psi_R^{\dagger \bar \alpha \bar
i}(z_4)>.\end{equation}  We suppress spin and flavour indices in
what follows, but it must be understood that we are dealing with
tensors throughout.   The first step is to regard the right-moving
fields as reflected left-moving ones: \begin{equation} \psi_R(z) \to
\psi_L(z^*).\end{equation}  Then we use Eq. (\ref{bosonizek}) to
express $G$ as a product of charge, spin and flavour Green's
functions: \begin{equation} G = G_cG_sG_f.\end{equation} The form
of $G_c$ is unique up to a multiplicative constant, and is
unaffected by the boundary. $G_s$ and $G_f$ are partially determined
by the bulk conformal field theories.  i.e. they must be solutions
of the linear Knizhnik-Zamolodchikov (KZ)
equations.\cite{Knizhnik}  These equations have two solutions in
both cases.  We label them $G_s(p)$ and $G_f(q)$ where $p$ and $q$
take values 0 and 1.  They are tensors in spin and flavour space;
we suppress these indices.  Thus we may schematically write the
solution as : \begin{equation}
G=\sum_{p,q=0}^1a_{p,q}G_cG_s(p)G_f(q).\label{4ptgen}\end{equation}
The four constants, $a_{p,q}$ remain to be determined.  It turns
out that three of them can be determined from general
considerations, with only one ($a_{1,1}$) depending on the boundary
conditions.

\begin{figure}
\epsfxsize=10 cm
\centerline{\epsffile{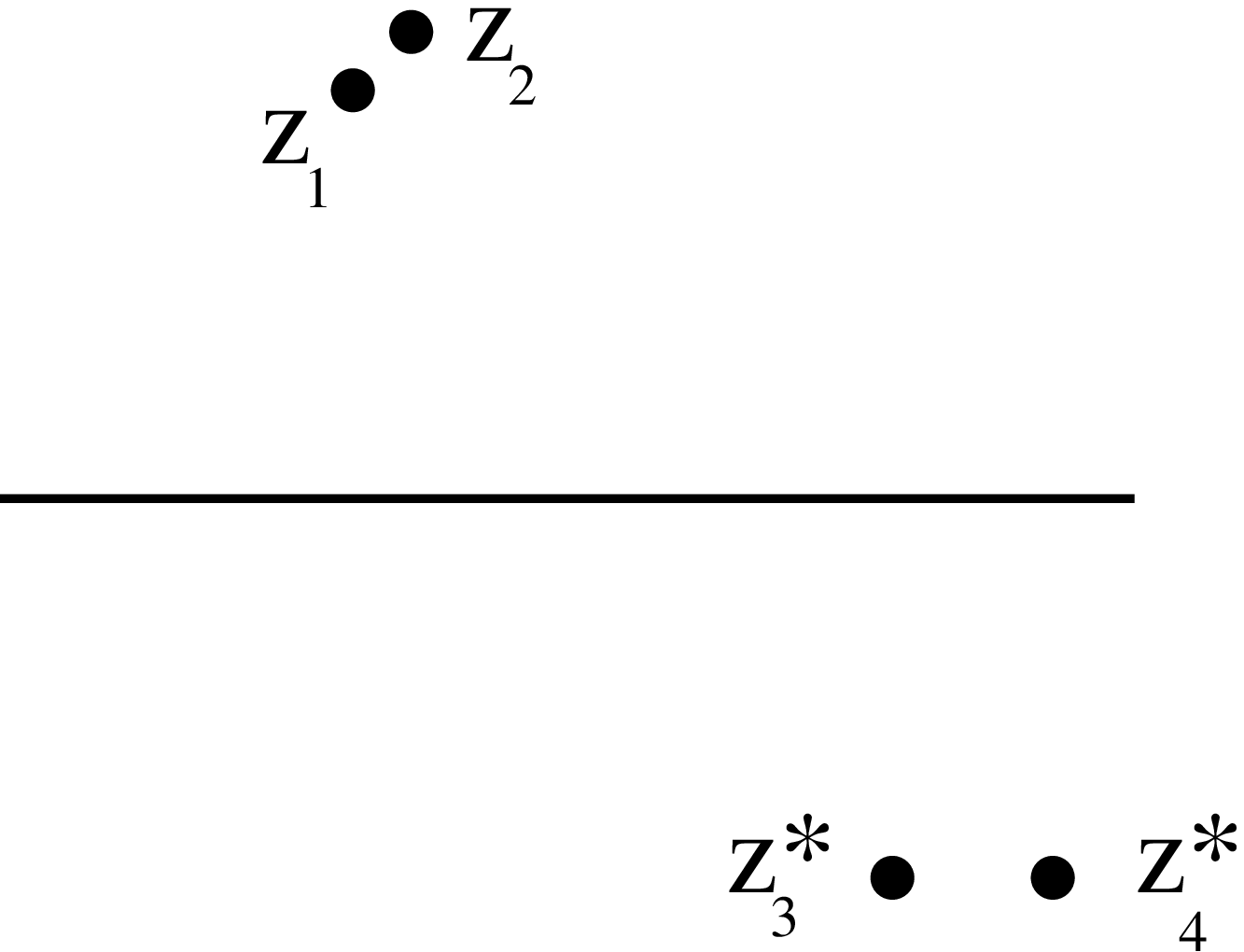}}
\caption{The bulk limit, $z_1\to z_2$, $z_3\to z_4$.}
\label{fig:4ptbulk}
\end{figure}

To see this, and to understand better how the solutions $G_s(p)$ and
$G_f(q)$ are defined, it is convenient to consider the limit $z_1\to
z_2$ and $z_3\to z_4$, illustrated in Figure (\ref{fig:4ptbulk}).
In the limit we may use the operator product expansion of $\psi_L
(z_1)$ with $\psi^\dagger_L(z_2)$.  Since these points are at a
fixed distance from the boundary as they approach each other, the
bulk OPE applies.  This is simply the trivial OPE of free fermions:
\begin{equation} \psi (z_1)\psi^\dagger (z_2) \sim {1\over z_1-z_2}
+ J  + \vec J + J^A + {\cal O}_s^{ad}{\cal O}^{ad}_f +
...\end{equation} Here we have used the fact that the spin and
flavour currents are bilinear in the fermion fields.  ${\cal
O}_s^{ad}$ and ${\cal O}^{ad}_f$ are the primary fields in the
adjoint representation of the spin and flavour groups.  These have
scaling dimension $2/(2+k)$ and $k/(2+k)$ respectively, so their
product has scaling dimension $1$ and corresponds to the product of
fermion fields with spin and flavour traces subtracted.  Note,
importantly, that neither adjoint primary can appear by itself,
since it has a fractional scaling dimension that does not occur in
the free fermion OPE.

Now let us consider the bosonized expression for the fermion
fields, of Eq. (\ref{bosonizek}).  Consider the OPE of each factor
separately: \begin{eqnarray} e^{i\sqrt{2\pi
/k}\phi}(z_1)e^{-i\sqrt{2\pi /k}\phi}(z_2) &\sim&
(z_1-z_2)^{-1/2k}+...\nonumber \\ g(z_1) g^\dagger (z_2) &\sim&
(z_1-z_2)^{-3/[2(2+k)]}+(z_1-z_2)^{1/[2(2+k)]}{\cal O}_s^{ad}
+...\nonumber \\ h(z_1)h^\dagger (z_2) &\sim &
(z_1-z_2)^{-(k^2-1)/[k(2+k)]}+ (z_1-z_2)^{1/[k(2+k)]}{\cal
O}_f^{ad}+...\end{eqnarray} A general solution of the KZ equations
for the four-point spin and flavour Green's functions, $G_s$ and
$G_f$ will have singularities corresponding to both the singlet and
adjoint terms in the OPE shown above.  The two independent
solutions, $G_s(p)$ and $G_f(q)$ can be chosen so that only the
singlet appears for $p=0$ and $q=0$ and only the adjoint appears
for $p=1$ and $q=1$.  It is now clear that, in order for $G$ to
reproduce free fermion behaviour in the limit $z_1\to z_2$, we must
demand that the coefficients $a_{1,0}$ and $a_{0,1}$ vanish in Eq.
(\ref{4ptgen}) so that the spin or flavour adjoint field doesn't
occur without being multiplied by the other adjoint field.
Similarly, requiring the correct normalization for the
singlet-singlet singularity in this limit: \begin{equation} G \to
{1\over (z_1-z_2)} {1\over (z_3^*-z_4^*)},\end{equation} determines
the non-zero value of the coefficient $a_{0,0}$.  i.e. it has the
same value as in the free fermion case, independent of the boundary
conditions. Thus only $a_{1,1}$ remains to be determined by detailed
consideration of the boundary condition.

{}From considering the limit $z_1\to z_2$, $z_3\to z_4$, we see
that: \begin{equation} a_{1,1} \propto <[{\cal O}^{ad}_s{\cal
O}^{ad}_f](z_1) [{\cal O}^{ad}_s{\cal
O}^{ad}_f](z_3^*)>_A.\end{equation} This is a two-point Green's
function with the two points staddling the boundary.  Thus its
normalization {\it does} depend on the particular boundary
condition, A.  In fact this normalization is precisely the
coefficient $C_A$ which was calcuated in the previous sub-section,
Eq. (\ref{C_Afusion}).  Thus we obtain,  upon identifying $\phi$ in
Eq. (\ref{C_Afusion}) with the adjoint (spin 1) primary in the spin
sector, for the Kondo boundary condition: \begin{equation} a_{1,1}
= {S^1_s/S^1_0\over
S^0_s/S^0_0}a_{1,1}^{\hbox{free}}.\end{equation} This result
completes the determination of the four point function.  The
general form of $G$ is rather complicated, involving hypergeometric
functions.  We only give explicit results here for the simplest
non-trivial case, $k=2$, $s=1/2$.  We consider only the most
singular ($2k_F$ part) of the spin density Green's function.
Writing:\begin{equation} {\cal S}^a(\vec r,\tau ) \equiv
e^{2ik_Fr}\psi_L^{\dagger}{\sigma^a\over 2}\psi_R +
h.c.,\end{equation} this becomes: \begin{equation} <S^a(\vec
r_1,\tau_1 )S^b(\vec r_2,\tau_2)> = {\delta^{ab}\over
8\pi^4r^2_1r_2^2}{\eta^{-1/2}\over |z_1-z_2^*|^2}\left[ 2\cos
2k_F(r_1+r_2) + \left(2+{\eta \over 1-\eta }\right)\cos
2k_F(r_1-r_2)\right]+ ...\end{equation} Here the $...$ represents
less singular terms and $\eta$ is the cross-ratio: \begin{equation}
\eta \equiv -{(z_1-z_1^*)(z_2-z_2^*)\over
(z_1-z_2^*)(z_2-z_1^*)}={4r_1r_2\over
(\tau_1-\tau_2)^2+(r_1+r_2)^2}.\end{equation}

\begin{figure}
\epsfxsize=10 cm
\centerline{\epsffile{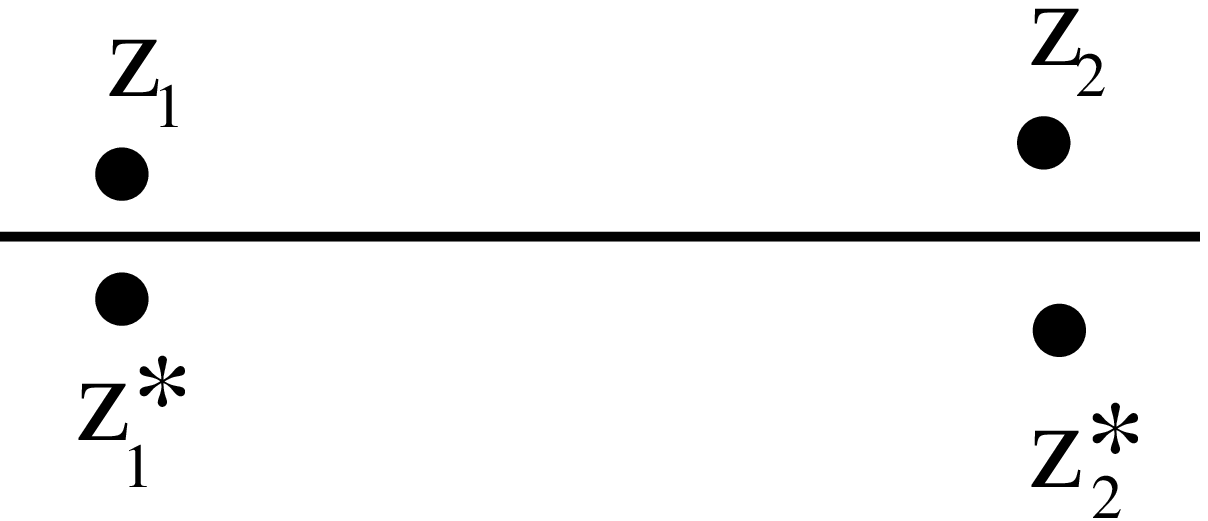}}
\caption{The boundary limit $z_1\to z_1^*$, $z_2\to
z_2^*$.}
\label{fig:4ptbound}
\end{figure}

Note that in the bulk limit, $|z_1-z_2|<<r_1,r_2$, [Figure
(\ref{fig:4ptbulk}) with $z_1=z_3$ and $z_2=z_4$] $\eta \to 1$ and
the Green's function reduces to its bulk form: \begin{equation}
<S^a(\vec r_1,\tau_1 )S^b(\vec r_2,\tau_2)>\to {\delta^{ab}\over
8\pi^4r^2_1r_2^2}{\cos 2k_F(r_1-r_2)\over
(\tau_1-\tau_2)^2+(r_1-r_2)^2},\end{equation} the free fermion
result.  On the other hand, in the boundary limit, [Figure
(\ref{fig:4ptbound})] $r_1,r_2<<|\tau_1-\tau_2|$, $\eta \to 0$ and
\begin{equation} <S^a(\vec r_1,\tau_1 )S^b(\vec r_2,\tau_2)>\to
{\delta^{ab}\cos 2k_Fr_1\cos 2k_Fr_2\over
4\pi^4(r_1r_2)^{3/2}|\tau_1-\tau_2|}.\end{equation} The
$\tau$-dependence implies that $S^a$ has scaling dimension 1/2 at
the boundary, although it has dimension 1 in the bulk,
corresponding to the spin current operator.

This change at the boundary can be understood as follows.  As
explained in the previous sub-section, in the presence of a
boundary, the spin density becomes a bilocal operator, at the
points $z$ and $z^*$.  As $r\to 0$ we use the O.P.E. to express it
as a sum of local operators.  In general, any operator in the
O.P.E. of the two left-moving operators ${\cal O}_L$ and $\bar
{\cal O}_L$ may appear. The coefficients of the various terms in
the O.P.E. (some of which may be zero) depend on the particular
boundary condition.  This is very natural since we are taking an
O.P.E. for two points which straddle the boundary.  [See Figure
(\ref{fig:1to2}).]  In this case we are obtaining the spin adjoint
primary field in the O.P.E., unaccompanied by the flavour adjoint,
as it is in the bulk O.P.E.  We give a general prescription for
determining the boundary operator content in the next sub-section.
\subsection{Boundary Operator Content and Leading Irrelevant
Operator: Specific Heat, Susceptibility, Wilson Ratio, Resistivity
at $T>0$} As we saw in Secs. I and III, the leading irrelevant
operator plays a very important role in the Kondo problem,
determining the low temperature behaviour of the specific heat,
susceptibility and resistivity.  One of the novel features of a
non-Fermi liquid boundary condition is that boundary operators may
appear which do not occur in the bulk theory.  As explained in the
previous sub-sections, this is a consequence of the bilocal nature
of  operators in the presence of a boundary and the fact that the
O.P.E. coefficients depend on the particular boundary condition.
In this sub-section we derive a general formula which gives all
boundary operators that occur with a particular boundary
condition.  Then we analyse the effect of the leading irrelevant
operator in the overscreened Kondo problem.

\begin{figure}
\epsfxsize=10 cm
\centerline{\epsffile{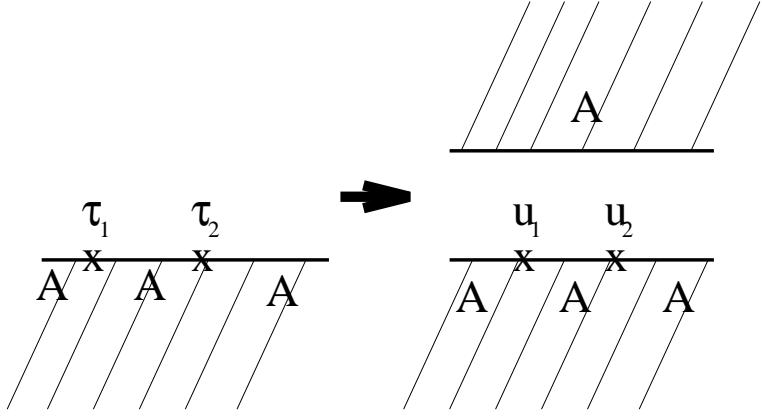}}
\caption{Conformal mapping of a boundary two-point
function from the semi-infinite plane to the finite
strip.}
\label{fig:plstrip}
\end{figure}

We can identify the boundary operator content from a general
relationship between the finite-size spectrum and operator content.
This is established by a conformal mapping from the semi-infinite
plane to the infinite strip.  We consider a correlation function for
some primary operator, ${\cal O}$, on the semi-infinite plane:
\begin{equation} <{\cal O}(\tau_1){\cal O}^\dagger
(\tau_2)>_A.\end{equation} [See Figure (\ref{fig:plstrip}).]  Now
we make the conformal mapping: \begin{equation} z = le^{\pi
w/l}.\end{equation}  Here $w$ is on the strip: \begin{equation}
-l/2< \hbox {Im}w <l/2.\end{equation}  We define: \begin{eqnarray}
z &=& \tau + ix \nonumber \\ w&=&u+iv.\end{eqnarray} It is
convenient to assume $\tau_1, \tau_2 >0$ so that $v_1=v_2=0$, as
shown in Figure (\ref{fig:plstrip}).  The correlation function on
the infinite plane has the form: \begin{equation} <{\cal
O}(\tau_1){\cal O}^\dagger (\tau_2)>_A={1\over
(\tau_1-\tau_2)^{2x}}.\end{equation}  From this we obtain the
correlation function on the strip: \begin{eqnarray} <{\cal
O}(u_1){\cal O}^\dagger (u_2)>_{AA} &=&\left\{{ {\partial z\over
\partial w}(u_1){\partial z\over \partial w}(u_2)\over
[z(u_1)-z(u_2)]^2}\right\}^x\nonumber \\ &=& \left[{2l\over
\pi}\sinh {\pi \over 2l}(u_1-u_2)\right]^{2x}.\end{eqnarray}  Here
$AA$ denotes the correlation function on the strip with boundary
condition $A$ on both sides.  As $u_2-u_1\to \infty$ this
approaches \begin{equation} <{\cal O}(\tau_1){\cal O}^\dagger
(\tau_2)>_A \to \left({\pi \over l}\right)^{2x}e^{-\pi
x(u_2-u_1)/l}.\end{equation}  Alternatively, we may evaluate the
correlation function on the strip by inserting a complete set of
states: \begin{equation} <{\cal O}(u_1){\cal O}^\dagger (u_2)>_{AA}=
\sum_n|<0|{\cal O}|n>_{AA}|^2e^{-E_n(u_2-u_1)}.\end{equation}  Here
we get the eigenstates on the strip with boundary conditions of
type `$A$' on both sides.  As $u_2-u_1\to \infty$, the lowest
energy state created from the groundstate by ${\cal O}$ dominates.
This is simply the highest weight state corresponding to the
primary field ${\cal O}$.  Clearly, the corresponding energy is:
\begin{equation} E_n = \pi x/l.\end{equation}  Thus we see that the
primary boundary operators with boundary condition $A$, are in
one-to correspondance with the conformal towers appearing in the
spectrum with boundary conditions $AA$.

Actually we have made an assumption here, that the operator ${\cal
O}$ does not change the boundary condition.  This appears to be
reasonable for the Kondo problem, but in some situations, such as
the X-ray edge singularity, it is neccessary to consider more
general boundary condition changing operators.\cite{Affleck10}

Thus to obtain the boundary operator content with a boundary
condition $A$, we simply need to know the finite size spectrum with
the boundary condition $A$ at both sides.  In the case of the Kondo
boundary condition, this is obtained by ``double fusion''.  i.e.
beginning with the free fermion spectrum, fusing once with the
spin-$s$ primary gives the spectrum with free fermion boundary
conditions at one end and Kondo boundary conditions at the other.
Fusing a second time with the spin-$s$ primary gives the spectrum
with Kondo boundary conditions at both ends, as mentioned in Sec.
III.  The fact that double fusion gives the spectrum with Kondo
boundary conditions at both ends can also be seen by a ``completing
the square'' argument.  i.e., if we have Kondo impurities at $x=0$
and $x=l$, then at a special value of both Kondo couplings,
$\lambda_1=\lambda_2=2/3$, the Hamiltonian returns to its free form
if we redefine the $SU(2)$ currents by: \begin{equation} \vec {\cal
J}_n \equiv \vec J_n + \vec S_1+(-1)^n\vec S_2.\end{equation}
Double fusion represents a generalization of ordinary angular
momentum addition rules, for the case of two impurity spins, which
is consistent with the conformal tower structure of the WZW model.
With free fermion boundary conditions, the boundary operators are
simply the bulk free fermion operators.  After double fusion, new
boundary operators, not present in the free fermion theory, are
generally present.  However, an exception occurs in the case $s=k/2$
corresonding to exact or underscreening.  In this case under double
fusion the jth conformal tower is mapped into itself:
\begin{equation} j  \to k/2-j \to j.\end{equation}  Thus the free
fermion operator content is recovered, corresponding to a Fermi
liquid boundary condition.

On the other hand, in the overscreened case, $k>2s$, non-trivial
operators are always obtained.  In particular, the spin $j=1$,
charge $Q=0$ flavour singlet operator, $\vec \phi$, always occurs.
This follows from double fusion with the identity operator:
\begin{eqnarray} (Q=0, j=0,\hbox{flavour singlet}) &\to & (Q=0,
j=s,\hbox{flavour singlet}) \nonumber \\ &\to& (Q=0,
j=0,1,...\hbox{min}(2s,k-2s),\hbox{flavour singlet}).\end{eqnarray}
For overscreening, min$(2s, k-2s)\geq 1$, so $\vec \phi$ (with
$j=1$) always occurs.  $\vec \phi$ itself, cannot appear in the
effective Hamiltonian, since it is not s spin singlet.  However, its
first descendent, $\vec J_{-1}\cdot \vec \phi$ {\it can} appear
since it is invariant under all the symmetries of the underlying
Kondo Hamiltonian.  It has dimension, \begin{equation} x= 1 +
{2\over 2+k}.\end{equation}  This dimension is $>1$ meaning that
the operator is irrelevant.  On the other hand, the dimension is
$<2$ so it is more relevant than $\vec {\cal J}^2$, the leading
irrelevant operator in the Fermi liquid case.  It can be seen to be
the leading irrelevant operator in all overscreened
cases.\cite{Affleck3}

We now discuss the calculation of various quantities to lowest
non-vanishing order in the leading irrelevant term in the effective
Hamiltonian: \begin{equation} \delta H = \lambda_1 \vec J_{-1}\cdot
\vec \phi .\end{equation} We begin with the specific heat.  The
simple change of variables trick that we used in Sec. III for the
Fermi liquid case doesn't work here.  In fact, since $\vec
J_{-1}\cdot \vec \phi$ is a Virasoro primary operator (although a
Kac-Moody descendant)  its one-point function continues to vanish
even at finite temperature.  Consequently, the leading contribution
to the specific heat is second order in $\lambda_1$.  The
contribution to the impurity free energy of $O(\lambda_1^2)$ is:
\begin{equation} -\beta f_{\hbox{imp}} = {\lambda_1^2\over
2}\int_{-\beta /2}^{\beta /2}d\tau_1\int_{-\beta /2}^{\beta
/2}d\tau_2T<\vec J_{-1}\cdot \vec \phi (\tau_1)\vec J_{-1}\cdot
\vec \phi (\tau_2)>.\label{cimp}\end{equation}  Since  $\vec
J_{-1}\cdot \vec \phi$ is a Virasoro primary, this two-point
function is given by: \begin{eqnarray} <\vec J_{-1}\cdot \vec \phi
(\tau_1)\vec J_{-1}\cdot \vec \phi (\tau_2)>&=&{3(2+k/2)\over
|\tau_1-\tau_2|^{2(1+\Delta )}}\nonumber \\ &\to &{3(2+k/2)\over
|{\beta \over \pi}\sin {\pi \over
\beta}(\tau_1-\tau_2)|^{2(1+\Delta )}}. \end{eqnarray}  Here
\begin{equation} \Delta \equiv {2\over 2+k},\end{equation} and we
have assumed that $\vec \phi$ has unit normalization:
\begin{equation} <\phi^a(\tau_1)\phi^b(\tau_2)> = {\delta^{ab}\over
|\tau_1-\tau_2|^{2\Delta}}.\end{equation} The normalization of
$\vec J_{-1}\cdot \vec \phi$ can then be fixed using the Kac-Moody
algebra.  A naive rescaling argument  implies that:
\begin{equation} f_{\hbox{imp}} \propto
T^{1+2\Delta}.\end{equation}  Actually the integral in Eq.
(\ref{cimp}) requires an ultraviolet cut-off.  However, this only
leads to additional terms in $f_{\hbox{imp}}$ that vanish more
rapidly as $T\to 0$.  Evaluating the universal term in Eq.
(\ref{cimp}) explicitly, we obtain: \begin{eqnarray}
C_{\hbox{imp}}&=&-T{\partial ^2f_{\hbox{imp}}\over \partial T^2}
=\lambda_1^2\left[T^{2\Delta}\pi^{1+2\Delta}(2\Delta)^23({k\over
2}+2) {1\over 2}{\Gamma (1/2-\Delta )\Gamma (1/2)\over \Gamma
(1-\Delta )}+ O(T)\right] + O(\lambda_1^3T^{3\Delta}) + ...
\nonumber \\ &&(\hbox{for}\  k>2)\end{eqnarray}  Here $\Gamma (x)$
is Euler's Gamma function.   Note that this result is independent
of the magnitude of the impurity spin, $s$.  Also note that since,
for $k>2$,  $\Delta =2/(2+k)<1/2$, so the impurity specific heat is
more singular than in the Fermi liquid case.  In the particular
case $k=2$, where $\Delta =1/2$, we obtain from Eq. (\ref{cimp}):
\begin{equation} C_{\hbox{imp}}=\lambda_1^2\pi^29T\ln (T_K/T) +
...\end{equation}  Note that this is only more singular by a log
than in the Fermi liquid case.

The impurity susceptibility is also second order in $\lambda_1$,
\begin{equation} \chi_{\hbox{imp}}\propto \beta \lambda_1^2\int
d\tau_1d\tau_2dx_1dx_2 <\vec J_{-1}\cdot \vec \phi (\tau_1) \vec
J_{-1}\cdot \vec \phi
(\tau_2)J^3(0,x_1)J^3(0,x_2)>.\label{chiimp}\end{equation}  Again,
by a scaling argument we see that: \begin{equation}
\chi_{\hbox{imp}} \propto \lambda_1^2T^{2\Delta -1}.\end{equation}
By evaluating the integral in Eq. (\ref{chiimp}) explicitly, we can
caluculate the Wilson ratio, $R_W$ for the overscreened Kondo
problem: \begin{equation} {\chi_{\hbox{imp}}\over
C_{\hbox{imp}}}{C_{\hbox{bulk}}\over \chi_{\hbox{bulk}}}\equiv R_W
= {(2+k/2)(2+k)^2\over 18}.\end{equation}

We can calculate the Wilson ratio in the exactly screened, Fermi
liquid, case, much more simply.  In this case, the simple rescaling
argument of Sec. III always works giving: \begin{equation}
{C_{\hbox{imp}}/C_s\over \chi_{\hbox{imp}}/\chi}=1.\end{equation}
Here $C_s$ is the part of the bulk specific heat coming from the
spin degrees of freedom.  Hence the Wilson ratio is given by
\begin{equation} R_W={C\over C_s}.\end{equation}  It simply
measures the ratio of the total bulk specific heat to that coming
{}from the spin degrees of freedom.  Using: \begin{equation} C = {\pi
\over 3}cT,\end{equation} where $c$ is the conformal anomaly, and
the fact that $c=2k$ for 2k species of free fermions and
$c=3k/(2+k)$ for the level k SU(2) WZW model, we obtain:
\begin{equation} R_W={2(2+k)\over 3},\end{equation} for the exactly
screened Kondo problem.

Finally we consider the T-dependence of the impurity resistivity.
For more details, see Ref. (\onlinecite{Affleck6}).  We may write
the electron Green's function in the one-dimensional theory with the
impurity at the origin in the form: \begin{equation}
G(r_1,-r_2,\omega_n)-G_0(r_1+r_2,\omega_n)=G_0(r_1,\omega_n)
\Sigma_1(\omega_n)G_0(r_2,\omega_n).\end{equation}  In the three
dimensional theory with a dilute random array of Kondo impurities,
of density $n_i$ the self-energy becomes: \begin{equation}
\Sigma_3={n_i\over \nu}\Sigma_1.\end{equation}  The electron
life-time is given by: \begin{equation} {1\over \tau}=-2\hbox{Im}
\Sigma_3^R,\end{equation} where the superscript $R$ denotes the
retarded Green's function, obtained by analytic continuation from
the Matsubara Green's function.  The conductivity is then expressed
in terms of the lifetime in the standard way: \begin{equation}
\sigma = {2e^2\over 3m^2}\int{d^3p\over (2\pi )^3}\left[{-dn_F\over
d\epsilon_p}\right]\vec p^2\tau (\epsilon_p).\end{equation}  Thus,
we just need to calculate the one-dimensional self-energy,
$\Sigma_1$.  This was already done, at T=0, in Subsection (VIC).
There we showed that: \begin{equation} G=G_0S^{(1)},\end{equation}
where $S^{(1)}$ is the scattering matrix element for 1-electron to
go into 1-electron, given explicitly in Eq. (\ref{S(1)}).  This
gave the resistivity: \begin{equation} \rho = {1-S^{(1)}\over
2}\rho_{\hbox{u}},\end{equation} where $\rho_{\hbox{u}}$ is the
unitary limit resistivity.  To obtain the leading T-dependence, we
do perturbation theory in the leading irrelevant coupling constant,
$\lambda_1$.  In this case, there is a non-zero first order
contribution: \begin{equation} \delta \Sigma_1 \propto \lambda_1
<\psi \vec J_{-1}\cdot \vec \phi \psi^\dagger >.\end{equation}
Since $\lambda_1$ has scaling dimension $\Delta$, it follows that:
\begin{equation} \rho = \rho_{\hbox{u}}{1-S^{(1)}\over 2}\left[1+
\alpha \lambda_1 T^{\Delta}+...\right],\end{equation} where $\alpha
$ is a dimensionless constant which can be obtained by explicit
evaluation of first order perturbation theory.  After a rather long
calculation, $\alpha$ can be expressed in terms of an integral over
hypergeometric functions.  We evaluated this integral numerically
for the case $k=2$, where $\Delta = 1/2$, obtaining:
\begin{equation} \alpha =4\sqrt{\pi}.\end{equation}  Thus we can
form another universal ratio involving the square of the
temperature-dependent term in the resistivity, and the specific
heat coefficient.

\begin{figure}
\epsfxsize=10 cm
\centerline{\epsffile{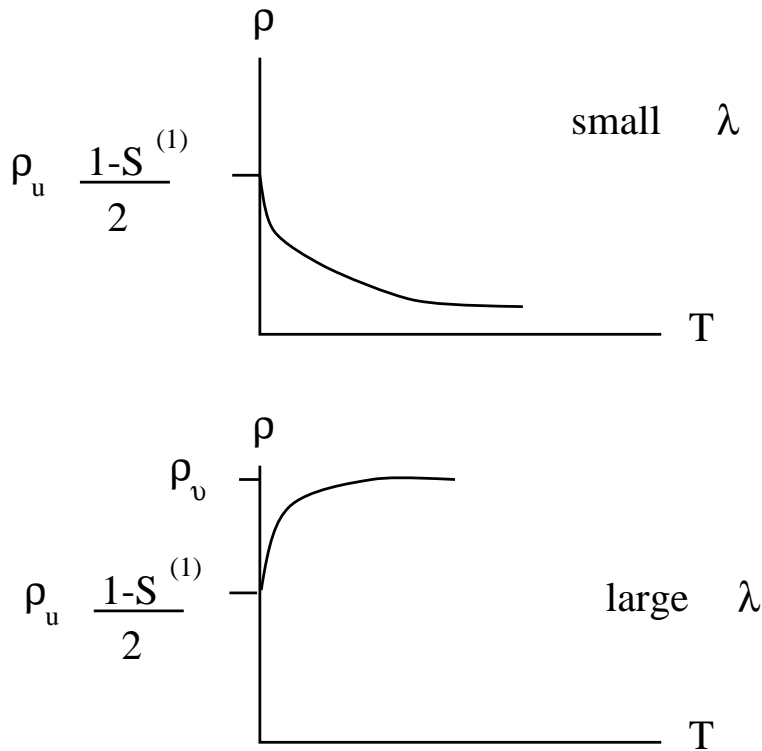}}
\caption{Qualitative behaviour of the resistivity in the
overscreened case.} \label{fig:resov}
\end{figure}

Note that in the Fermi liquid case the resistivity had the form:
\begin{equation} \rho = \rho_{\hbox{u}}[1-\alpha
\lambda_1^2T^2].\end{equation} The leading T-dependence is second
order in $\lambda_1$ and $T$.  In the non-Fermi liquid case the
sign of the leading T-dependence depends on the sign of the
irrelevant coupling constant, $\lambda_1$.  As we increase the
Kondo coupling constant, so as to pass through the non-trivial
critical point, at Kondo couplings of $O(1)$, the leading
irrelevant coupling constant, $\lambda_1$, should change sign.
Thus the sign of the leading T-dependent term in the resistivity
actually switches at the critical point.  We can't determine the
value of $\lambda_1$ from our methods.  However, if we make the
plausible assumption that $\lambda_1 <0$ for weak Kondo coupling
and hence $\lambda_1 >0$ for strong Kondo coupling, then the
resistivity behaves, at low T as shown in Figure (\ref{fig:resov})
(since the constant, $\alpha >0$). For weak Kondo coupling the
resistivity decreases with T, exhibiting a power law singularity at
$T=0$.  On the other hand, for strong Kondo coupling, the
resistivity initially {\it increases} with $T$ [Figure
(\ref{fig:resov})].  This is quite reasonable for very strong Kondo
coupling where, at high $T$ the effective renormalized Kondo
coupling is close to the unstable fixed point at $\infty$.  Hence,
we expect to obtain the unitary limit resistivity at high $T$ in
this case.  This requires $\rho$ to increase with $T$.

This $\sqrt{T}$ behaviour of the resistivity, and other scaling
behaviour predicted by the present approach for the case $k=2$,
$s=1/2$, were  observed in recent experiments.\cite{Ralph1}
However, the interpretation of these experiments in terms of the
2-channel Kondo problem is controversial at
present.\cite{Wingreen,Ralph2,Moustakas}

\acknowledgements I would especially like to thank my main
collaborator in this work, Andreas Ludwig.  I would also like to
thank Chandra Varma for interesting me in this problem, Nathan
Seiberg for suggesting the ``fusion rule hypothesis'' and John
Cardy and Dan Cox  for their collaboration and helpful
suggestions.  I am very grateful to Ming Yu who produced a \TeX
version, from my transparencies, of some lectures  which I gave in
Beijing in June, 1991. This essentially became  Sections I-III.
This research was supported in part by NSERC of Canada.

 \end{document}